\documentclass{article}%
\usepackage{amsmath}
\usepackage{graphicx}
\usepackage{amsfonts}
\usepackage{amssymb}%
\providecommand{\U}[1]{\protect\rule{.1in}{.1in}}
\providecommand{\U}[1]{\protect\rule{.1in}{.1in}}
\providecommand{\U}[1]{\protect\rule{.1in}{.1in}}
\providecommand{\U}[1]{\protect\rule{.1in}{.1in}}
\providecommand{\U}[1]{\protect\rule{.1in}{.1in}}
\newtheorem{theorem}{Theorem}
\newtheorem{acknowledgement}[theorem]{Acknowledgement}

\begin{document}

\title{{\Large Self-adjoint extensions and spectral analysis in the generalized
Kratzer problem}}
\author{M.C. Baldiotti\thanks{Institute of Physics, University of Sao Paulo, Brazil;
e-mail: baldiott@fma.if.usp.br}, D.M. Gitman\thanks{Institute of Physics,
University of Sao Paulo, Brazil; e-mail: gitman@dfn.if.usp.br}, I.V.
Tyutin\thanks{Lebedev Physical Institute, Moscow, Russia; e-mail:
tyutin@lpi.ru}, and B.L. Voronov\thanks{Lebedev Physical Institute, Moscow,
Russia; e-mail: voronov@lpi.ru}}
\date{}
\maketitle

\begin{abstract}
We present a mathematically rigorous quantum-mechanical treatment of a
one-dimensional nonrelativistic motion of a particle in the potential field
\[
V(x)=g_{1}x^{-1}+g_{2}x^{-2},\ x\in\mathbb{R}_{+}=\left[  0,\infty\right)  ~.
\]
For $g_{2}>0$ and $g_{1}<0$, the potential is known as the \textit{Kratzer
potential} $V_{K}(x)$ and is usually used to describe molecular energy and
structure, interactions between different molecules, and interactions between
non-bonded atoms.

We construct all self-adjoint Schr\"{o}dinger operators with the potential
$V(x)$ and represent rigorous solutions of the corresponding spectral
problems. Solving the first part of the problem, we use a method of specifying
s.a. extensions by (asymptotic) s.a. boundary conditions. Solving spectral
problems, we follow the Krein's method of guiding functionals. This work is a
continuation of our previous works devoted to Coulomb, Calogero, and
Aharonov-Bohm potentials.

\end{abstract}

\section{Introduction\label{S1}}

In this article, we present a mathematically rigorous quantum-mechanical (QM)
treatment of a one-dimensional nonrelativistic motion on a semiaxis of a
spinless particle of mass $m$ in the potential field
\begin{equation}
V(x)=g_{1}x^{-1}+g_{2}x^{-2},\ x\in\mathbb{R}_{+}=\left[  0,\infty\right)  ~.
\label{pot}%
\end{equation}
On the physical level of rigor, the Schr\"{o}dinger equation with potential
(\ref{pot}) was studied for a long time in connection with different physical
problems, see for example \cite{Fue26,Scar58} and books \cite{Flu94,LanLi77}.
In particular, this potential enters the stationary radial Schr\"{o}dinger
equation
\begin{equation}
\left[  \frac{d^{2}}{dr^{2}}+\frac{2m}{\hbar^{2}}\left(  E_{nl}-U\left(
r\right)  -\frac{l\left(  l+1\right)  \hbar^{2}}{2mr^{2}}\right)  \right]
\psi_{nl}\left(  r\right)  =0~, \label{radial}%
\end{equation}
where $n$ and $l$ are radial and angular quantum numbers, after separating
spherical variables in three-dimensional spherically symmetric QM problems,
see e.g. \cite{LanLi77}. The potential (\ref{pot}) is singular at the origin,
it is repulsive at this point for $g_{2}>0$, and has a minimum at a point
$x_{0}>0$ for $g_{2}>0$ and $g_{1}<0$. The potential with\textrm{ }%
$g_{1},g_{2}$ in the latter range\textit{ }is known as the \textit{Kratzer
potential} \cite{Kra20}. The Kratzer potential is conventionally used to
describe molecular energy and structure, interactions between different
molecules \cite{BayBoC07}, and interactions between nonbonded atoms
\cite{Bau53}. For $g_{2}<0$ and $g_{1}>0$, we have the \textit{inverse Kratzer
potential }which is conventionally used \ to describe tunnel effects,
scattering of charged particles \cite{Mas49} and decays, in particular,
molecule ionization and fluorescence \cite{BalMi90}. In addition, valence
electrons in a hydrogen-like atom are described in terms of such a potential
\cite{Eli62}. When modeling some physical systems, a constant is usually added
to the angular momentum term, $l\left(  l+1\right)  \rightarrow\beta+l\left(
l+1\right)  $, in order to take some effective potential energy into account.
For example, in the model of a molecule interaction, $\beta$ can represent the
dissociation energy of a diatomic molecule\ \cite{BayBoC07} or, in the
scattering problem, this parameter represents attractive ($\beta<0$)\ or
repulsive ($\beta>0$)\ interactions between charged particles \cite{Mas49}.

In Figure 1 we show the shape of the potential under consideration for
different values of the parameters.%

\begin{figure}
[ptb]
\begin{center}
\includegraphics[
height=1.9168in,
width=3.3439in
]%
{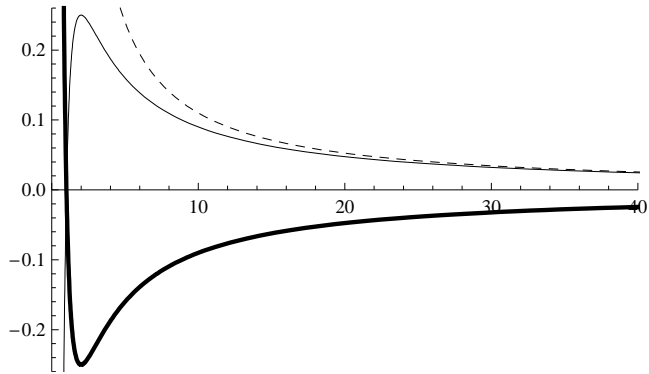}%
\caption{Potential $V(x)=g_{1}x^{-1}+g_{2}x^{-2}$, with $g_{1}=g_{2}=1$
(dashed), $g_{1}=-g_{2}=1$ (solid) and $g_{1}=-g_{2}=-1$ (thick).}%
\end{center}
\end{figure}

Even though a number of works was devoted to the QM problem with the potential
(\ref{pot}), a rigorous mathematical analysis of this problem is lacking in
the literature. The aim of such an analysis (which is, in fact, the aim of the
present article) is to construct all self-adjoint (s.a. in what follows)
Schr\"{o}dinger operators (Hamiltonians) with the potential (\ref{pot}) and
present rigorous solutions of the corresponding spectral problems.

When solving the first part of the problem, we use a method for specifying
s.a.\textrm{ }differential operators by (asymptotic) s.a. boundary conditions
(the so-called alternative method, see \cite{VorGiT107}). When solving
spectral problems, we follow the Krein's method of guiding functionals, see
\cite{Krein46} and books \cite{Naima69}. This work is a continuation of our
previous works \cite{VorGiT307,GitTySV09} devoted to Coulomb, Calogero, and
Aharonov-Bohm potentials; using the given references, the reader can become
acquainted with necessary basic notions and constructions, like guiding
functional and Green function.

As in the above-mentioned works, we start with a s.a. differential operation
$\check{H}$ on $\mathbb{R}_{+}$,%

\begin{equation}
\check{H}=-d_{x}^{2}+g_{1}x^{-1}+g_{2}x^{-2},\label{7.3.0}%
\end{equation}
and examining solutions of the corresponding homogeneous differential equation
$(\check{H}-W)\psi=0$, or%
\begin{equation}
\psi^{\prime\prime}-(g_{1}x^{-1}+g_{2}x^{-2}-W)\psi=0,\;W=|W|\mathrm{e}%
^{i\varphi},\;0\leq\varphi<2\pi~,\label{7.3.1}%
\end{equation}
which is the Schr\"{o}dinger equation (with omitted factor $2m/\hbar^{2}$)
with a complex energy $W$, for $\operatorname{Im}W=0$, we write $W=E$ in what follows.

The basic operator $\hat{H}^{+}$ in $L^{2}\left(  \mathbb{R}_{+}\right)  $
associated with $\check{H}$ is defined on the natural domain\footnote{a.c.
means absolutely continuous.} $D_{\check{H}}^{\ast}(\mathbb{R}_{+})\subset
L^{2}\left(  \mathbb{R}_{+}\right)  $,
\begin{equation}
D_{\check{H}}^{\ast}\mathrm{\ }(\mathbb{R}_{+})=\left\{  \psi_{\ast}%
(x):\psi_{\ast},\psi_{\ast}^{\prime}\,\mathrm{are\,a.c.\;in\;}\mathbb{R}%
_{+};\,\psi_{\ast},\check{H}\psi_{\ast}\in L^{2}\left(  \mathbb{R}_{+}\right)
\right\}  ~,\label{1}%
\end{equation}
it is the adjoint of the so-called initial symmetric operator $\hat{H}$
associated with $\check{H}$ and defined on the dense domain $D_{H}%
=\mathcal{D}\left(  \mathbb{R}_{+}\right)  $, the space of smooth
functions\ with a compact support,%
\begin{equation}
\mathcal{D}\left(  \mathbb{R}_{+}\right)  =\left\{  \psi\left(  x\right)
:\psi\in C^{\infty}(\mathbb{R}_{+}),~\mathrm{supp}\psi\subseteq\lbrack
\alpha,\beta]\subset(0,\infty)\right\}  \label{novo1*}%
\end{equation}
it is evident that $\mathcal{D}\left(  \mathbb{R}_{+}\right)  \subset
D_{\check{H}}^{\ast}(\mathbb{R}_{+})$ and $\hat{H}\subset\hat{H}^{+}$.\textrm{
}The operator $\hat{H}^{+}$is generally not self-adjoint and even not
symmetric; its quadratic asymmetry form is denoted by $\Delta_{H^{+}}$. All
possible Hamiltonians associated with $\check{H}$ are defined as s.a.
restrictions of $\hat{H}^{+}$, which simultaneously are s.a. extensions of the
symmetric $\hat{H}$, the restrictions to some subspaces (domains) belonging to
$D_{\check{H}}^{\ast}\mathrm{\ }(\mathbb{R}_{+})$ and specified by some
additional (asymptotic) s.a. boundary conditions on functions belonging to
$D_{\check{H}}^{\ast}\mathrm{\ }(\mathbb{R}_{+})$ under which the asymmetry
form $\Delta_{H^{+}}$ becomes trivial (vanishes); these domains are maximum
subspaces in $D_{\check{H}}^{\ast}\mathrm{\ }(\mathbb{R}_{+})$ where the
operator $\hat{H}^{+}$ is symmetric\footnote{Although the notions
\textquotedblleft s.a. extension of $\hat{H}^{+}$\textquotedblright\ and
\textquotedblleft s.a. restriction of $\hat{H}$ \textquotedblright\ are
equivalent; it is more customary to speak about s.a. extensions; we use one or
another of the equivalent notions where appropriate.} (see \cite{VorGiT107}).
Our first aim is to describe all these Hamiltonians. The special case of
$g_{1}=0$ corresponds to the Calogero potential and was already considered in
\cite{VorGiT307}, we therefore keep $g_{1}\neq0$ in what follows.

This paper is organized as follows. In sec. \ref{S2} we present and discuss
some exact solutions of equation (\ref{7.3.1}) and their asymptotics. In the
following five sections, we construct all s.a. extensions of $\hat{H}$, and
perform the corresponding spectral analysis of the Hamiltonians for different
ranges of the parameter\textrm{ }$g_{2}$. In secs. (\ref{SS3.1}-\ref{SS3.4}),
we consider the case of $g_{2}\neq0$. The special case of $g_{2}=0$ is
considered in sec. \ref{SS3.5}. In sec. \ref{S4}, we highlight some remarks
and possible applications of the obtained results.

\section{Exact solutions and asymptotics\label{S2}}

We first consider the Schr\"{o}dinger equation (\ref{7.3.1}). Introducing a
new variable $z$ and new functions $\phi_{\pm}(z)$ instead of the respective
$x$ and $\psi\left(  x\right)  $,
\begin{align}
&  z=\lambda x,\ \lambda=2\sqrt{-W}=2\sqrt{|W|}\mathrm{e}^{i(\varphi-\pi
)/2},\;\psi(x)=x^{1/2\pm\mu}\mathrm{e}^{-z/2}\phi_{\pm}(z)~,\nonumber\\
&  \mu=\left\{
\begin{array}
[c]{l}%
\sqrt{g_{2}+1/4},\ g_{2}\geq-1/4\\
i\varkappa,\ \varkappa=\sqrt{|g_{2}|-1/4},\ g_{2}<-1/4
\end{array}
\right.  , \label{7.3b.1}%
\end{align}
we reduce eq. (\ref{7.3.1}) to the confluent hypergeometric equations for
$\phi_{\pm}(z)$,
\begin{align}
&  zd_{z}^{2}\phi_{\pm}(z)+(\beta_{\pm}-z)d_{z}\phi_{\pm}(z)-\alpha_{\pm}%
\phi_{\pm}(z)=0~,\nonumber\\
&  \alpha_{\pm}=1/2\pm\mu+g_{1}/\lambda,\ \beta_{\pm}=1\pm2\mu\ ,
\label{7.3b.2}%
\end{align}
their solutions are the known confluent hypergeometric functions $\Phi
(\alpha_{\pm},\beta_{\pm};z)\,$and $\Psi(\alpha_{\pm},\beta_{\pm};z)$, see
\cite{BatEr53,GraRy71}.

Solutions $\psi\left(  x\right)  $ of eq. (\ref{7.3.1}) are restored from
solutions of eqs. (\ref{7.3b.2}) by transformation (\ref{7.3b.1}). In what
follows, we use $u_{1}(x;W)$, $u_{2}(x;W)$, and $\upsilon_{1}\left(
x;W\right)  $ defined by
\begin{align}
&  u_{1}\left(  x;W\right)  =x^{1/2+\mu}\mathrm{e}^{-z/2}\Phi(\alpha_{+}%
,\beta_{+};z)=\left.  u_{1}\left(  x;W\right)  \right\vert _{\lambda
\rightarrow-\lambda}\ ,\nonumber\\
&  u_{2}\left(  x;W\right)  =x^{1/2-\mu}\mathrm{e}^{-z/2}\Phi(\alpha_{-}%
,\beta_{-};z)=\left.  u_{2}\left(  x;W\right)  \right\vert _{\lambda
\rightarrow-\lambda}=\left.  u_{1}\left(  x;W\right)  \right\vert
_{\mu\rightarrow-\mu}\ ,\nonumber\\
&  \upsilon_{1}\left(  x;W\right)  =\lambda^{2\mu}x^{1/2+\mu}\mathrm{e}%
^{-z/2}\Psi(\alpha_{+},\beta_{+};z)=\lambda^{2\mu}\frac{\Gamma(-2\mu)}%
{\Gamma(\alpha_{-})}u_{1}+\frac{\Gamma(2\mu)}{\Gamma(\alpha_{+})}u_{2}\ .
\label{7.3b.3}%
\end{align}
The function $u_{2}\,$is\ not defined for $\beta_{-}=-n$, or $\mu
=(n+1)/2,n\in\mathbb{Z}_{+}$, in particular, for $\mu=1/2$. For such $\mu$, we
replace $u_{2}$ by other solutions of eq. (\ref{7.3.1}), they are considered
in the subsequent sections.

The coefficients of the Taylor expansion of functions $u_{1}(x;W)/x^{1/2+\mu}$
and $u_{2}(x;W)/x^{1/2-\mu}$ with respect to $x$ are polynomials in $\lambda$.
Because these functions are even in $\lambda$, the coefficients are
polynomials in $W$, whence it follows that $u_{1}\left(  x;W\right)  $ and
$u_{2}\left(  x;W\right)  $ are entire functions in $W$ at any point $x$
except $x=0$ for $u_{2}$ with $\mu>1/2$.

If $g_{2}\geq-1/4$ ($\mu\geq0$), then $u_{1}\left(  x;W\right)  $ and
$u_{2}\left(  x;W\right)  $ are real-entire functions of $W$. If $g_{2}<-1/4$
($\mu=i\varkappa$), then $u_{2}\left(  x;E\right)  =\overline{u_{1}\left(
x;E\right)  }$.

The pairs $u_{1},u_{2}$ with $\mu\neq0\,$and $u_{1},\upsilon_{1}$ for
$\operatorname{Im}W\neq0$ are the fundamental systems of solutions of eq.
(\ref{7.3.1}) because the respective Wronskians are
\begin{equation}
\mathrm{Wr}\left(  u_{1},u_{2}\right)  =-2\mu,\;\mathrm{Wr}\left(
u_{1},\upsilon_{1}\right)  =-\Gamma(\beta_{+})/\Gamma(\alpha_{+})\equiv
-\omega(W)~. \label{7.3b.4a}%
\end{equation}

The well-known asymptotics of the special functions $\Phi$ and $\Psi$, see
e.g. \cite{BatEr53}, entering solutions (\ref{7.3b.3}) allows simply
estimating \ the asymptotic behavior of the solutions at the origin, as
$x\rightarrow0$, and at infinity, as $x\rightarrow\infty$.

As $x\rightarrow0$, we have%
\begin{align}
&  u_{1}(x;W)=\kappa_{0}^{-1/2-\mu}u_{1\mathrm{as}}(x)+O(x^{3/2+\mu
}),\ \nonumber\\
&  u_{2}(x;W)=\kappa_{0}^{-1/2+\mu}u_{2\mathrm{as}}(x)+\left\{
\begin{array}
[c]{l}%
O(x^{5/2-\mu}),\;-1/4<g_{2}<3/4,\ g_{2}\neq0,\\
(0<\mu<1,\mu\neq1/2)\\
O(x^{3/2}),\;g_{2}<-1/4\;(\mu=i\varkappa)
\end{array}
\right.  ,\label{7.3.1a}%
\end{align}
and, if $\alpha_{+}\neq-n$, $\alpha_{-}\neq-m$,$\,\,n,m\in\mathbb{Z}_{+}$,%
\begin{equation}
\upsilon_{1}(x;W)=\left\{
\begin{array}
[c]{l}%
\frac{\Gamma(2\mu)}{\Gamma(\alpha_{+})}x^{1/2-\mu}(1+O(x)),\ \ g_{2}%
\geq3/4\ (\mu\geq1)\\
\lambda^{2\mu}\frac{\Gamma(-2\mu)}{\Gamma(\alpha_{-})}\kappa_{0}^{-1/2-\mu
}u_{1\mathrm{as}}(x)+\frac{\Gamma(2\mu)}{\Gamma(\alpha_{+})}\kappa
_{0}^{-1/2+\mu}u_{2\mathrm{as}}(x)+O(x^{3/2}),\\
-1/4<g_{2}<3/4,g_{2}\neq0\ (0<\mu<1,\mu\neq1/2)\\
\lambda^{2i\varkappa}\frac{\Gamma(-2i\varkappa)}{\Gamma(\alpha_{-})}\kappa
_{0}^{-1/2-i\varkappa}u_{1\mathrm{as}}(x)+\frac{\Gamma(2i\varkappa)}%
{\Gamma(\alpha_{+})}\kappa_{0}^{-1/2+i\varkappa}u_{2\mathrm{as}}%
(x)+O(x^{3/2}),\\
\ g_{2}<-1/4\ (\mu=i\varkappa)
\end{array}
\right.  ,\label{7.3b.4}%
\end{equation}
where%
\begin{subequations}
\begin{align}
u_{1\mathrm{as}}(x) &  =(\kappa_{0}x)^{1/2+\mu},\nonumber\\
u_{2\mathrm{as}}(x) &  =\left\{
\begin{array}
[c]{l}%
(\kappa_{0}x)^{1/2-\mu}-\frac{g_{1}/\kappa_{0}}{2\mu-1}(\kappa_{0}x)^{3/2-\mu
},\ -1/4<g_{2}<3/4,\ g_{2}\neq0,\\
(0<\mu<1,\mu\neq1/2)\\
(\kappa_{0}x)^{1/2-i\varkappa},\ g_{2}<-1/4\ (\mu=i\varkappa)
\end{array}
\right.  ,\label{7.3.1b}%
\end{align}
and $\kappa_{0}$ is an arbitrary, but fixed, parameter of dimension of inverse length.

As $x\rightarrow\infty,\ \operatorname{Im}W>0$, we have%
\end{subequations}
\begin{align*}
&  u_{1}(x;W)=\frac{\Gamma(\beta_{+})}{\Gamma(\alpha_{+})}\lambda^{\alpha
_{+}-\beta_{+}}x^{g_{1}/\lambda}\mathrm{e}^{z/2}(1+O(x^{-1}))=O(x^{a}%
\mathrm{e}^{|W|^{1/2}\sin(\varphi/2)})~,\\
&  \upsilon_{1}(x;W)=\lambda^{-\alpha_{-}}x^{-g_{1}/\lambda}\mathrm{e}%
^{-z/2}(1+O(x^{-1}))=O(x^{-a}\mathrm{e}^{-|W|^{1/2}\sin(\varphi/2)})~,\\
&  a=2^{-1}|W|^{-1/2}g_{1}\sin(\varphi/2)~.
\end{align*}

The obtained asymptotics are sufficient to allow definite conclusions about
the deficiency indices of the initial symmetric operator $\hat{H}$ as
functions of the parameters $g_{1},g_{2}$ and thereby about a possible variety
of its s.a. extensions. It is evident that for $\operatorname{Im}W>0$ the
function $u_{1}(x;W)$ exponentially increasing at infinity and is not
square-integrable. The function $\upsilon_{1}\left(  x;W\right)  $
exponentially decreasing at infinity is not square-integrable at the origin
for $g_{2}\geq3/4$ ($\mu\geq1)$, whereas for $g_{2}<3/4$, it is (moreover, for
$g_{2}<3/4$, any solution of eq. (\ref{7.3.1}) is square-integrable at the
origin). Because for $\operatorname{Im}W>0$, the functions $u_{1},\upsilon
_{1}$ form a fundamental system of eq. (\ref{7.3.1}), this equation with
$\operatorname{Im}W>0$ has no square-integrable solutions for $g_{2}\geq3/4$,
whereas for $g_{2}<3/4$, there exists one square-integrable solution,
$\upsilon_{1}\left(  x;W\right)  $. This means that the deficiency indices of
the initial symmetric operator $\hat{H}$ are equal to zero, $m_{\pm}=0$, for
$g_{2}\geq3/4$ and are equal to unity, $m_{\pm}=1$, for $g_{2}<3/4$.

Correspondingly for $g_{2}\geq3/4$, there is a unique s.a. extension of
$\hat{H}$, whereas for $g_{2}<3/4$, there exists a one-parameter family of
s.a. extensions of $\hat{H}$. A structure of these extensions, in particular,
an appearance of their specifying asymptotic boundary conditions, depends
crucially on a specific range of values of the parameter $g_{2}$. In what
follows, we distinguish five such regions and consider them separately.

\section{Self-adjoint extensions and spectral analysis\label{S3}}

\subsection{The first range $g_{2}\geq3/4$ $\left(  \mu\geq1\right)
$\label{SS3.1}}

As was mentioned above, the deficiency indices of the initial symmetric
operator $\hat{H}$ with $g_{2}\,$in this range are zero. This implies that for
$g_{2}\geq3/4$, the operator $\hat{H}^{+}$ is s.a. and $\hat{H}_{1}=\hat
{H}^{+}$\ is a unique s.a. extension of $\hat{H}$ with the domain
$D_{H_{{\large \mathfrak{e}}}}=D_{\check{H}}^{\ast}\left(  \mathbb{R}%
_{+}\right)  $ (\ref{1}).

A spectral analysis of the \ s.a. operator $\hat{H}_{1}=\hat{H}^{+}$ begins
with an evaluation of its Green function $G\left(  x,y;W\right)  $ that is the
kernel of the integral representation of the solution $\psi_{\ast}\left(
x\right)  $ of the inhomogeneous differential equation%
\begin{equation}
\left(  \check{H}-W\right)  \psi_{\ast}\left(  x\right)  =\eta(x),\,\eta(x)\in
L^{2}(\mathbb{R}_{+})~ \label{natDom}%
\end{equation}
\textrm{ }with $\operatorname{Im}W\neq0$\textrm{ }under the condition that
$\psi_{\ast}\in$ $D_{\check{H}}^{\ast}\left(  \mathbb{R}_{+}\right)  $, i.e.,
that $\psi_{\ast}$ is square-integrable\footnote{We note, that $D_{\check{H}%
}^{\ast}\left(  \mathbb{R}_{+}\right)  $ can be considered as the space of
unique square-integrable solutions of eq. (\ref{natDom}) with
$\operatorname{Im}W\neq0$ and any $\,\eta(x)\in L^{2}(\mathbb{R}_{+}).$},
$\psi_{\ast}(x)\in L^{2}(\mathbb{R}_{+})$ (see \cite{VorGiT307,GitTySV09}%
).\ The general solution of this equation without the condition of square
integrability can be represented as%
\begin{align}
&  \psi_{\ast}(x)=a_{1}u_{1}(x;W)+a_{2}\upsilon_{1}(x;W)+I(x;W)~,\nonumber\\
&  \psi_{\ast}^{\prime}(x)=a_{1}u_{1}^{\prime}(x;W)+a_{2}\upsilon_{1}^{\prime
}(x;W)+I^{\prime}(x;W)~, \label{7.3.3a}%
\end{align}
where

\textrm{ }%
\begin{align*}
&  I(x;W)=\int_{0}^{x}G^{\left(  +\right)  }\left(  x,y;W\right)
\eta(y)dy+\int_{x}^{\infty}G^{\left(  -\right)  }\left(  x,y;W\right)
\eta(y)dy~,\\
&  I^{\prime}(x;W)=\int_{0}^{x}d_{x}G^{\left(  +\right)  }\left(
x,y;W\right)  \eta(y)dy+\int_{x}^{\infty}d_{x}G^{\left(  -\right)  }\left(
x,y;W\right)  \eta(y)dy~,\\
&  G^{\left(  +\right)  }\left(  x,y;W\right)  =\omega^{-1}(W)\upsilon
_{1}(x;W)u_{1}(y;W)~,\\
&  G^{\left(  -\right)  }\left(  x,y;W\right)  =\omega^{-1}(W)u_{1}%
(x;W)\upsilon_{1}(y;W)~,
\end{align*}
with $\omega$ given in (\ref{7.3b.4a}). Using the \ Cauchy-Bunyakovskii
inequality, it is easy to show that $I(x;W)$ is bounded as $x\rightarrow
\infty$. The condition $\psi_{\ast}(x)\in L^{2}(\mathbb{R}_{+})$ then implies
that $a_{1}=0$, because $u_{1}(x;W)$ exponentially grows while $\upsilon
_{1}(x;W)$ exponentially decreases at infinity. As $x\rightarrow0$, we have
$I(x)\sim O(x^{3/2})$, $I^{\prime}(x)\sim O(x^{1/2})$ (up to the logarithmic
accuracy at $g_{2}=3/4$), whereas $\upsilon_{1}(x;W)$ is not square-integrable
at the origin. The condition $\psi_{\ast}(x)\in L^{2}(\mathbb{R}_{+})$ then
implies that $a_{2}=0$. In addition, we see that the asymptotic behavior of
functions $\psi_{\ast}(x)$ belonging to $D_{\check{H}}^{\ast}\left(
\mathbb{R}_{+}\right)  $ at the origin, as $x\rightarrow0$, is estimated by%
\begin{equation}
\psi_{\ast}(x)=O(x^{3/2}),\ \psi_{\ast}^{\prime}(x)=O(x^{1/2}%
)~.\label{natdom2}%
\end{equation}
\ Together with the fact that the functions $\psi_{\ast}$ vanish at infinity
(see below), this implies that the asymmetry form $\Delta_{H^{+}}$ is trivial,
which confirms that in the first range the operator $\hat{H}^{+}$ is symmetric
and therefore self-adjoint (in contrast to the next ranges considered in the
subsequent sections).

It follows that the Green's function is given by
\[
G\left(  x,y;W\right)  =\left\{
\begin{array}
[c]{c}%
G^{\left(  +\right)  }\left(  x,y;W\right)  ,\;x>y\\
G^{\left(  -\right)  }\left(  x,y;W\right)  ,\;x<y
\end{array}
\right.  ~.
\]

The representation (\ref{7.3b.3}) of the function $\upsilon_{1}$ in terms of
the functions $u_{1}$ and $u_{2}$ is inconvenient sometimes, because the
individual summands do not exist for some $\mu$ although $\upsilon_{1}$ does.
For our purposes, another representations are convenient.\textrm{ }For
$m-1<2\mu<m+1$, $m\geq2$, the function $\upsilon_{1}(x;W)$ can be represented
as%
\begin{align*}
&  \upsilon_{1}(x;W)=A_{m}(W)u_{1}(x;W)+\frac{\omega(W)}{2\mu}\upsilon
_{(m)}(x;W)~,\\
&  A_{m}(W)=\lambda^{2\mu}\frac{\Gamma(-2\mu)}{\Gamma(\alpha_{-})}%
+a_{m}(W)\frac{\Gamma(2\mu)\Gamma(\beta_{-})}{\Gamma(\alpha_{+})}~,\\
&  \upsilon_{(m)}(x;W)=u_{2}\left(  x;W\right)  -a_{m}(W)\Gamma(\beta
_{-})u_{1}\left(  x;W\right)  ~,\\
&  a_{m}(W)=\lambda^{m}\frac{\Gamma(\alpha_{+m})}{m!\Gamma(\alpha_{-m}%
)},\;\alpha_{\pm m}=\frac{1\pm m}{2}+g_{1}/\lambda~.
\end{align*}
It is easy to see that all the coefficients $a_{m}(W)$ are polynomials in $W$
which are real for $\operatorname{Im}W=0$ ($W=E$). In view of the relation%

\[
\lim_{\beta\rightarrow-n}^{-1}\Gamma(\beta)\Phi(\alpha,\beta;x)=\frac
{x^{n+1}\Gamma(\alpha+n+1)}{(n+1)!\Gamma(\alpha)}\Phi(\alpha+n+1,n+2;x)
\]
(see \cite{GraRy71,BatEr53}), the functions $\upsilon_{(m)}(x;W)$ and
$A_{m}(W)$ exist for $m-1<2\mu<m+1$ and for any $W$. In fact, $\upsilon
_{(m)}(x;W)$ are particular solutions of eq. (\ref{7.3.1}) which are
real-entire in $W$ and have the properties (for $m-1<2\mu<m+1$)%
\[
\mathrm{Wr}(u_{1},\upsilon_{(m)})=-2\mu,\;\upsilon_{(m)}(x;W)=x^{1/2-\mu
}(1+O(x)),\;x\rightarrow0~.
\]

As a guiding functional, we take%
\begin{equation}
\Phi(\xi;W)=\int_{0}^{\infty}U(x;W)\xi(x)dx,\;\xi\in\mathbb{D}=D_{r}%
(\mathbb{R}_{+})\cap D_{H_{\mathfrak{e}}}~,\label{A.1.1}%
\end{equation}
where $U\left(  x;W\right)  =u_{1}\left(  x;W\right)  $ and $D_{r}%
(\mathbb{R}_{+})$ is the space of arbitrary functions with a support bounded
from the right: $\varphi\left(  x\right)  \in D_{r}\left(  \mathbb{R}%
_{+}\right)  \Longrightarrow\mathrm{supp\,}\varphi\subseteq\left[
0,\beta\right]  $, $\beta<\infty$; the domain $\mathbb{D}$ is dense in
$L^{2}(\mathbb{R}_{+})$. The functional $\Phi(\xi;W)$ (\ref{A.1.1}) is a
simple guiding functional, i.e., it satisfies the properties: 1) for a fixed
$\xi$, the functional $\Phi(\xi;W)$ is an entire function of $W$; 2) if
$\Phi(\xi_{0};E_{0})=0$,$\;\operatorname{Im}E_{0}=0$,$\;\xi_{0}\in\mathbb{D}$,
then the inhomogeneous\ equation $(\check{H}-E_{0})\psi=\xi_{0}$ has a
solution $\psi\in\mathbb{D}$; 3) $\Phi(\check{H}\xi;W)=W\Phi(\xi;W)$. It is
easy to verify the properties 1) and 3), and it remains to verify that the
property 2) also holds. Let%
\begin{equation}
\Phi(\xi_{0};E_{0})=\int_{0}^{b}u_{1}(x;E_{0})\xi_{0}(x)dx=0,\;\xi_{0}%
\in\mathbb{D},\;\mathrm{supp}\xi_{0}\in\lbrack0,b]~.\label{A.1.2}%
\end{equation}
We consider the function $\psi(x)$ defined by%
\begin{equation}
\psi(x)=\frac{1}{2\mu}\left[  u_{1}(x;E_{0})\int_{x}^{b}\upsilon_{(m)}%
(y;E_{0})\xi_{0}(y)dy+\upsilon_{(m)}(x;E_{0})\int_{0}^{x}u_{1}(y;E_{0})\xi
_{0}(y)dy\right]  \label{A.1.3}%
\end{equation}
that evidently satisfying the equation $(\check{H}-E_{0})\psi(x)=\xi_{0}(x).$
Using condition (\ref{A.1.2}), we obtain that $\mathrm{supp}\psi\in
\lbrack0,b]$, i.e., $\psi\in D_{r}(\mathbb{R}_{+})$, and therefore, $\psi\in
L^{2}(c,b)$ for any $c>0$. With taking \ the asymptotic behavior of functions
$u_{1}(x;E_{0})$, $\upsilon_{(m)}(x;E_{0})$, and $\xi_{0}(x)$ at the origin
into account, a simple evaluation of the integrals in representation
(\ref{A.1.3}) gives:%
\[
\psi(x)=\left\{
\begin{array}
[c]{l}%
O(x^{1/2+\mu}),\;1\leq\mu<3\\
O(x^{7/2}\ln\delta),\;\mu=3\\
O(x^{7/2}),\;\mu>3
\end{array}
\right.  ,\;x\rightarrow0~,
\]
i.e., $\psi\in D_{H_{\mathfrak{e}}}$, and therefore, $\psi\in\mathbb{D}$.

The derivative of the spectral function is given by%
\begin{equation}
\sigma^{\prime}(E)=\pi^{-1}\operatorname{Im}\left[  \omega^{-1}(E+i0)A_{m}%
(E+i0)\right]  ~. \label{7.3c.10b}%
\end{equation}

Because $\omega^{-1}(W)A_{m}(W)$ is an analytic function of $\mu$, its value
at $\mu=m/2$ is a limit as$\,\mu\rightarrow m/2$. For $\mu\neq m/2$,
representation (\ref{7.3c.10b}) can be simplified to%
\[
\sigma^{\prime}(E)=\operatorname{Im}\Omega(E+i0),\;\Omega(W)=\frac
{\lambda^{2\mu}\Gamma(-2\mu)\Gamma(\alpha_{+})}{\pi\Gamma(\alpha_{-}%
)\Gamma(\beta_{+})}~.
\]

For $E=p^{2}\geq0$, $p\geq0$, $\lambda=2p\mathrm{e}^{-i\pi/2}$, we find%
\begin{equation}
\sigma^{\prime}(E)=\left(  \frac{|\Gamma(\alpha_{+})|}{\Gamma(\beta_{+}%
)}\right)  ^{2}\frac{(2p)^{2\mu}\mathrm{e}^{-\pi g_{1}/2p}}{2\pi}>0~.
\label{7.3c.5}%
\end{equation}
We see that $\sigma^{\prime}(E)$ is a nonsingular function for $E\geq0$. It
follows that the spectrum of the s.a. Hamiltonian $\hat{H}_{1}$ is continuous
for all such values of $E$.

For $E=-\tau^{2}<0$, $\tau>0$, $\lambda=2\tau$, the function $\Omega(E)$ is
real for all values of $E$ where $\Omega(E)$ is finite, which implies that
$\operatorname{Im}\Omega(E+i0)$ can differ from zero only at the discrete
points $E_{n}$ where $1/$ $\Omega(E_{n})=0$. It is easy to see that the latter
equation is reduced to the equations $\alpha_{+}(E_{n})=-n,n\in\mathbb{Z}_{+}%
$, which have solutions only if $g_{1}<0$, and the solutions $E_{n}$ are then
given by%
\begin{equation}
E_{n}=-g_{1}^{2}(1+2\mu+2n)^{-2}~,\;\tau_{n}=\left\vert g_{1}\right\vert
\left(  1+2\mu+2n\right)  ^{-1}~. \label{7.3c.3}%
\end{equation}
We thus obtain that for $E<0$, the function $\sigma^{\prime}(E)\,$is equal to
zero if $\ g_{1}>0\ $, whereas if $g_{1}<0$, this function is given by%
\[
\sigma^{\prime}(E)=\sum_{n=0}^{\infty}Q_{n}^{2}\delta(E-E_{n})~,\;Q_{n}%
=\frac{(2\tau_{n})^{\mu+1}}{\Gamma(\beta_{+})}\sqrt{\frac{\tau_{n}%
\Gamma(1+2\mu+n)}{\left\vert g_{1}\right\vert n!}}~.
\]

The final result of this section is as follows.

For $g_{2}>3/4$ $(\mu>1)$, the spectrum of a unique s.a. operator
(Hamiltonian) $\hat{H}_{1}$ is simple and given by%
\[
\mathrm{spec}\hat{H}_{1}=\left\{
\begin{array}
[c]{l}%
\mathbb{R}_{+},\ g_{1}>0\\
\mathbb{R}_{+}\cup\{E_{n},\},\ g_{1}<0
\end{array}
\right.  .
\]

For $g_{1}>0$, its generalized eigenfunctions $U_{E}\left(  x\right)
=\sqrt{\sigma^{\prime}(E)}u_{1}(x;E)$,\ $E\geq0$, form a complete
orthonormalized system in $L^{2}(\mathbb{R}_{+})$. For $g_{1}<0$, the
generalized eigenfunctions $U_{E}\left(  x\right)  =\sqrt{\sigma^{\prime}%
(E)}u_{1}(x;E)$,\ $E\geq0$, of the continuous spectrum and the eigenfunctions
$U_{n}(x)=Q_{n}u_{1}(x;E_{n}),\ n\in\mathbb{Z}_{+}$, of the discrete spectrum
form a complete orthonormalized system in $L^{2}(\mathbb{R}_{+})$.

\subsection{The second range $3/4>g_{2}>-1/4,\ g_{2}\neq0\ (1>\mu>0,\ \mu
\neq1/2)$\label{SS3.2}}

We note that in this section, we consider the range $3/4>g_{2}>-1/4$ excluding
the point $g_{2}=0\,(\mu=1/2)$, the reason is that the function $u_{2}$ we use
here is not defined for $\mu=1/2$. The case $g_{2}=0\,(\mu=1/2)$ is considered
separately in the last subsection.

The operator $\hat{H}^{+}$ with $g_{2}$ in the second range is not s.a., and
we must construct its s.a. reductions. In accordance with the general
procedure of the alternative method, see \cite{VorGiT107} and also
\cite{VorGiT307}, \cite{GitTySV09} for examples, we begin with evaluating the
quadratic asymmetry form $\Delta_{H^{+}}$ in terms of quadratic boundary
forms, which are determined by the asymptotics of functions $\psi_{\ast}(x)$
belonging to the natural domain $D_{\check{H}}^{\ast}\left(  \mathbb{R}%
_{+}\right)  $ at the origin (the left boundary form) and at infinity (the
right boundary form). Because the potential vanishes at infinity, the right
boundary form is trivial (zero)\footnote{Moreover, we can prove that
$\psi_{\ast}$ vanishes at infinity together with its derivative, $\psi_{\ast
}(x),\psi_{\ast}^{\prime}(x)\overset{x\rightarrow\infty}{\longrightarrow}0$.},
see \cite{VorGiT107}, and the asymmetry form $\Delta_{H^{+}}$ is reduced to
(minus) the left boundary form. To determine an asymptotic behavior of
functions $\psi_{\ast}$ at the origin, we consider these functions as
solutions of the inhomogeneous eq. (\ref{natDom}) with $W=0$. Because in the
range under consideration, any solution of the homogeneous eq. (\ref{7.3.1})
is square-integrable at the origin, the general solution of eq. (\ref{natDom})
with $W=0$ can be represented as%
\begin{align}
&  \psi_{\ast}(x)=a_{1}u_{1}(x;0)+a_{2}u_{2}(x;0)\nonumber\\
&  -\frac{1}{2\mu}\int_{0}^{x}\left[  u_{1}(x;0)u_{2}(y;0)-u_{2}%
(x;0)u_{1}(y;0)\right]  \eta(y)dy~.\label{7.3.4d}%
\end{align}

The asymptotic behavior of the functions $u_{1}$ and $u_{2}$ in representation
(\ref{7.3.4d}) as $x\rightarrow0$ is given by (\ref{7.3.1a}) and
(\ref{7.3.1b}), the asymptotic behavior of the integral terms is estimated
using the Cauchy-Bunyakovskii inequality, and we find%
\begin{align}
&  \psi_{\ast}(x)=a_{1}u_{1\mathrm{as}}(x)+a_{2}u_{2\mathrm{as}}%
(x)+O(x^{3/2})~,\nonumber\\
&  \psi_{\ast}^{\prime}(x)=a_{1}u_{1\mathrm{as}}^{\prime}(x)+a_{2}%
u_{2\mathrm{as}}^{\prime}(x)+O(x^{1/2})~.\label{7.3.4e}%
\end{align}
With these asymptotics, we calculate the left boundary form $[\psi_{\ast}%
,\psi_{\ast}](0)=\lim_{x\rightarrow0}(-\overline{\psi}_{\ast}(x)\psi_{\ast
}^{\prime}(x)+\overline{\psi}_{\ast}^{\prime}(x)\psi_{\ast}(x))$ and obtain a
representation of the quadratic asymmetry form as a quadratic form in the
coefficients $a_{1}$ and $a_{2}\,$\ in (\ref{7.3.4e}):%
\[
\Delta_{H^{+}}(\psi_{\ast})=-2\mu k_{0}(\overline{a_{1}}a_{2}-\overline{a_{2}%
}a_{1})~.
\]
The coefficients $a_{1},a_{2}$ are called the (left) asymptotic boundary
(a.b.) coefficients\footnote{The inertia indices of the quadratic form
$(1/2i\mu\kappa_{0})\Delta_{+}$ are $1,1$, which confirms the previos
assertion in sec. (\ref{S2}) that the deficiency indices of $\hat{H}$ are
$m_{\pm}=1$, see \cite{VorGiT107}.}. The requirement on the a.b. coefficients
that $\Delta_{H^{+}}$vanish results in the relation\footnote{Here and in what
follows we use the notation $\mathbb{S}\left(  a,b\right)  =\left[
a,b\right]  ,\ a\sim b.$}%
\begin{equation}
a_{2}\sin\nu=a_{1}\cos\nu~,\;\nu\in\mathbb{S}\left(  -\pi/2,\pi/2\right)
~,\label{7.3.4f}%
\end{equation}
between these coefficients. It follows that the quadratic asymmetry form
$\Delta_{H^{+}}$ becomes trivial on the subspaces of $D_{\check{H}}^{\ast}$
such that the a.b. coefficients of functions $\psi_{\ast}(x)$ belonging to
$D_{\check{H}}^{\ast}$ satisfy relation (\ref{7.3.4f}) with fixed $\nu$. These
subspaces are just the domains of s.a. restrictions of $\hat{H}^{+}$, and
relation (\ref{7.3.4f}), with fixed $\nu$, defines the asymptotic boundary
conditions specifying these s.a. operators.

We thus obtain that for each $g_{2}$ in the second range, there exists a
family of s.a. Hamiltonians $\hat{H}_{2,\nu}$ parametrized by the parameter
$\nu$ on a circle with the domains $D_{H_{2\nu}}$ that are the subspaces of
functions belonging to $D_{\check{H}}^{\ast}(\mathbb{R}_{+})$ and having the
following asymptotic behavior at the origin, as $x\rightarrow0$,%
\begin{align}
&  \psi(x)=C\psi^{\mathrm{as}}(x)+O(x^{3/2})~,\;\psi^{\prime}(x)=C\psi
^{\mathrm{as}\prime}(x)+O(x^{1/2})~,\nonumber\\
&  \psi^{\mathrm{as}}(x)=u_{1\mathrm{as}}(k_{0}x)\sin\nu+u_{2\mathrm{as}%
}(x,k_{0})\cos\nu~. \label{7.3.6d}%
\end{align}

The spectral analysis of $\hat{H}_{2,\nu}$ is similar to that for $\hat{H}%
_{1}$ in the previous section, the difference is that the function
$\upsilon_{1}(x;W)$ is now square-integrable at the origin and we must take
asymptotic boundary conditions (\ref{7.3.6d}) into account. To evaluate the
Green's function for $\hat{H}_{2,\nu}$, we take the representation
(\ref{7.3.3a}) with $a_{1}=0$ for $\psi_{\ast}(x)$ belonging to $D_{H_{2\nu}}%
$, boundary conditions (\ref{7.3.6d}), and asymptotics (\ref{7.3.1a}),
(\ref{7.3.1b}) then yield%
\begin{align*}
&  a_{2}=k_{0}^{-2\mu}\omega^{-1}(W)\left[  \frac{\Gamma(2\mu)}{\Gamma
(\alpha_{+})}\sin\nu-\frac{\Gamma(-2\mu)(\lambda/k_{0})^{2\mu}}{\Gamma
(\alpha_{-})}\cos\nu\right]  ^{-1}\\
&  \times\cos\nu\int_{0}^{\infty}\upsilon_{1}(x;W)\eta(x)dx~.
\end{align*}
Representing the function $\upsilon_{1}(x;W)$ in the form%
\begin{align*}
&  \upsilon_{1}(x;W)=(2\mu)^{-1}k_{0}^{-1/2}\lambda^{\mu}[\tilde{\omega
}_{2,\nu}(W)u_{2,\nu}(x;W)+\omega_{2,\nu}(W)\tilde{u}_{2,\nu}(x;W)]~,\\
&  u_{2,\nu}(x;W)=k_{0}^{1/2+\mu}u_{1}(x;W)\sin\nu+k_{0}^{1/2-\mu}%
u_{2}(x;W)\cos\nu~,\\
&  \tilde{u}_{2,\nu}(x;W)=-k_{0}^{1/2+\mu}u_{1}(x;W)\cos\nu+k_{0}^{1/2-\mu
}u_{2}(x;W)\sin\nu~,\\
&  \omega_{2,\nu}(W)=\omega(W)(\lambda/k_{0})^{-\mu}\sin\nu+(\lambda
/k_{0})^{\mu}\frac{\Gamma(\beta_{-})}{\Gamma(\alpha_{-})}\cos\nu~,\\
&  \tilde{\omega}_{2,\nu}(W)=\omega(W)(\lambda/k_{0})^{-\mu}\cos\nu
-(\lambda/k_{0})^{\mu}\frac{\Gamma(\beta_{-})}{\Gamma(\alpha_{-})}\sin\nu~,
\end{align*}
where $\omega$ is given in (\ref{7.3b.4a}), the functions $u_{2,\nu}(x;W)$ and
$\tilde{u}_{2,\nu}(x;W)$ are real-entire in $W$ solutions of eq. (\ref{7.3.1})
and $u_{2,\nu}(x;W)$ satisfies boundary condition (\ref{7.3.6d}), we obtain
the Green function%
\begin{align}
&  G(x,y;W)=(2\mu k_{0})^{-1}\Omega(W)u_{2,\nu}(x;W)u_{2,\nu}(y;W)\nonumber\\
&  +\frac{1}{2\mu k_{0}}\left\{
\begin{array}
[c]{c}%
\tilde{u}_{2,\nu}(x;W)u_{2,\nu}(y;W),\ x>y\\
u_{2,\nu}(x;W)\tilde{u}_{2,\nu}(y;W),\ x<y
\end{array}
\right.  ~, \label{7.3.6s}%
\end{align}
where%
\begin{equation}
\Omega(W)=\omega_{2,\nu}^{-1}(W)\tilde{\omega}_{2,\nu}(W)~. \label{7.3.6t}%
\end{equation}
We note that the second summand in (\ref{7.3.6s}) is real for real $W=E$.

As a guiding functional we take the functional $\Phi(\xi;W)$ given by
(\ref{A.1.1}) with $U\left(  x;W\right)  =u_{2,\nu}(x;W)$ and $\xi
\in\mathbb{D}=D_{r}(\mathbb{R}_{+})\cap D_{H_{2,\nu}}$. The domain
$\mathbb{D}$ is dense in $L^{2}(\mathbb{R}_{+})$, $\overline{\mathbb{D}}%
=L^{2}(\mathbb{R}_{+})$. Following the procedure of the previous section, \ we
show that $\Phi(\xi;z)$ is a simple guiding functional, i.e., satisfies the
properties 1)-3) cited in subsec. \ref{SS3.1}. It is easy to verify the
properties 1) and 3). We prove that the property 2) also holds. Let%
\begin{equation}
\Phi(\xi_{0};E_{0})=\int_{0}^{b}u_{2,\nu}(x;E_{0})\xi_{0}(x)dx=0,\;\xi_{0}%
\in\mathbb{D},\;\mathrm{supp}\xi_{0}\in\lbrack0,b]~. \label{A.2.2}%
\end{equation}
We consider the function%
\[
\psi(x)=\frac{1}{2\mu\kappa_{0}}\left[  u_{2,\nu}(x;E_{0})\int_{x}^{b}%
\tilde{u}_{2,\nu}(y;E_{0})\xi_{0}(y)dy+\tilde{u}_{2,\nu}(x;E_{0})\int_{0}%
^{x}u_{2,\nu}(y;E_{0})\xi_{0}(y)dy\right]  ~,
\]
which is a solution of equation%
\[
(\check{H}-E_{0})\psi(x)=\xi_{0}(x)~.
\]
Using condition (\ref{A.2.2}), we obtain that $\mathrm{supp}\psi\in
\lbrack0,b]$, i.e., $\psi\in D_{r}(\mathbb{R}_{+})$, and therefore $\psi\in
L^{2}(c,b)$ for any $c>0$.

The function $\psi(x)$ allows the representation%
\begin{align}
&  \psi(x)=cu_{2,\nu}(x;E_{0})+\tilde{u}_{2,\nu}(x;E_{0})\int_{0}^{x}u_{2,\nu
}(y;E_{0})\xi_{0}(y)dy\nonumber\\
&  \,-u_{2,\nu}(x;E_{0})\int_{0}^{x}\tilde{u}_{2,\nu}(y;E_{0})\xi
_{0}(y)dy,\;c=\frac{1}{2\mu\kappa_{0}}\int_{0}^{b}\tilde{u}_{2,\nu}%
(y;E_{0})\xi_{0}(y)dy~.\label{A.2.3}%
\end{align}
Using the asymptotics of functions $u_{2,\nu}(x;E_{0})$, $\tilde{u}_{2,\nu
}(x;E_{0})$, and $\xi_{0}(x)$ and simple estimates of the asymptotic behavior
of the integral terms at the origin, we obtain that the asymptotic of
$\psi(x)$ at the origin is given by%
\[
\psi(x)=cu_{2,\nu}(x;E_{0})+O(x^{5/2-\mu}),\;x\rightarrow0~,
\]
which implies that $\psi\in D_{H_{\mathfrak{2,\nu}}}$ and therefore $\psi
\in\mathbb{D}$.

The derivative of the spectral function reads%
\[
\sigma^{\prime}(E)=\left(  2\pi\mu k_{0}\right)  ^{-1}\operatorname{Im}%
\Omega(E+i0)~.
\]

It is convenient to consider the cases $|\nu|<\pi/2$ and $\nu=\pm\pi/2$ separately.

We first consider the case $\nu=\pi/2$ where we have%
\begin{align*}
&  u_{2,\pi/2}(x;W)=k_{0}^{1/2+\mu}u_{1}(x;W),\\
&  \sigma^{\prime}(E)=\operatorname{Im}\Omega(E+i0),\ \Omega(W)=-\frac
{\Gamma(\beta_{-})\Gamma(\alpha_{+})(\lambda/k_{0})^{2\mu}}{2\pi\mu
k_{0}\Gamma(\beta_{+})\Gamma(\alpha_{-})}~.
\end{align*}

For $E=p^{2}\geq0$, $p\geq0$, $\lambda=2p\mathrm{e}^{-i\pi/2}$, we have%
\[
\sigma^{\prime}(E)=\frac{|\Gamma(\alpha_{+})|^{2}}{\Gamma^{2}(\beta_{+})}%
\frac{(2p/k_{0})^{2\mu}\mathrm{e}^{-\pi g_{1}/2p}}{2\pi k_{0}}~,
\]
such that $\sigma^{\prime}(E)$ is finite and $\mathrm{spec}H_{\mathfrak{2,\pi
/2}}=\mathbb{R}_{+}$.

For $E=-\tau^{2}<0$, $\tau>0$, $\lambda=2\tau$, the function $\Omega(E)$ is
real for all values of $E$ where $\Omega(E)$ is finite, which implies that
$\operatorname{Im}\Omega(E+i0)$ can differ from zero only at the discrete
points $E_{n}$ where $1/$ $\Omega(E_{n})=0$. The latter equation is reduced to
the equations $\alpha_{-}(E_{n})=-n,n\in\mathbb{Z}_{+}$, ($\Gamma(\alpha
_{+})=\infty$) or%
\begin{equation}
1+2\mu+g_{1}/\tau_{n}=-2n,\ n\in\mathbb{Z}_{+}~.\label{7.3.8}%
\end{equation}

Eqs. (\ref{7.3.8}) have no solutions for $g_{1}>0$ and for $g_{1}<0$ we have
(we will denote the points of discrete spectrum for $\nu=\pm\pi/2$ by
$\mathcal{E}_{n}$)%
\[
\tau_{n}=\frac{|g_{1}|}{1+2\mu+2n},\;\mathcal{E}_{n}=-\tau_{n}^{2}%
=-\frac{g_{1}^{2}}{(1+2\mu+2n)^{2}}~,
\]
such that we obtain%

\[
\sigma^{\prime}(E)=\sum_{n=0}^{\infty}Q_{n}^{2}\delta(E-\mathcal{E}%
_{n})~,\;Q_{n}=\frac{(2\tau_{n})^{\mu+1}k_{0}^{-(1/2+\mu)}}{\Gamma(\beta_{+}%
)}\sqrt{\frac{\Gamma(1+2\mu+n)}{\left(  1+2\mu+2n\right)  n!}}~.
\]

It is easy to see that for the case of $\nu=-\pi/2$, we obtain the same
results for spectrum and eigenfunctions as it must be\textrm{.}

The final result for the Hamiltonian $\hat{H}_{2,\pm\pi/2}$ is as follows. Its
spectrum is simple and given by%
\[
\mathrm{spec}\hat{H}_{2,\pm\pi/2}=\left\{
\begin{array}
[c]{l}%
\mathbb{R}_{+},\ g_{1}>0\ ,\\
\mathbb{R}_{+}\cup\{\mathcal{E}_{n},\ n\in\mathbb{Z}_{+}\},\ g_{1}<0
\end{array}
\right.  ,
\]
and the complete orthonormalized system of its eigenfunctions in
$L^{2}(\mathbb{R}_{+})$ is given by%
\begin{align*}
&  U_{E}\left(  x\right)  =\sqrt{\sigma^{\prime}(E)}k_{0}^{1/2+\mu}%
u_{1}(x;E),\ E\geq0~,\\
&  U_{n}(x)=\frac{2^{1-\mu}|\mathcal{E}_{n}|^{3/4-\mu/2}}{|g_{1}|^{1/2}%
|\Gamma(\beta_{-})|}\sqrt{q_{n}}u_{2}(x;\mathcal{E}_{n})~,\\
&  \ q_{n}=\left\{
\begin{array}
[c]{c}%
\Gamma^{-1}(1+n)\Gamma(1+n-2\mu),\;0<\mu<1/2\\
\Gamma^{-1}(2+n)\Gamma(2+n-2\mu),\;1/2<\mu<1
\end{array}
\right.  ,\;n\in\mathbb{Z}_{+}~,
\end{align*}
for $g_{1}>0$, and by%
\begin{align*}
&  U_{E}\left(  x\right)  =\sqrt{\sigma^{\prime}(E)}k_{0}^{1/2+\mu}%
u_{1}(x;E),\ E\geq0~,\\
&  U_{n}(x)=Q_{n}k_{0}^{1/2+\mu}u_{1}(x;\mathcal{E}_{n})~,
\end{align*}
for $g_{1}<0$.

Now, we turn to the case $|\nu|<\pi/2$. In this case we have%
\begin{align*}
&  \sigma^{\prime}(E)=\left(  2\pi\mu k_{0}\cos^{2}\nu\right)  ^{-1}%
\operatorname{Im}F_{2,\nu}^{-1}(E+i0)~,\\
&  F_{2,\nu}(W)=f_{2}(W)+\tan\nu,\;f_{2}(W)=\frac{\Gamma(\beta_{-}%
)\Gamma(\alpha_{+})(\lambda/k_{0})^{2\mu}}{\Gamma(\beta)\Gamma(\alpha_{-})}~.
\end{align*}

For $E=p^{2}\geq0$, $p\geq0$, $\lambda=2p\mathrm{e}^{-i\pi/2}$, we have%
\begin{equation}
\sigma^{\prime}(E)=\frac{B(E)}{2\pi k_{0}\cos^{2}\nu\lbrack A^{2}(E)+\mu
^{2}B^{2}(E)]}\ , \label{7.3.6a}%
\end{equation}
where $A(E)=\operatorname{Re}F_{2,\nu}(E)$ and $\mu B(E)=-\operatorname{Im}%
F_{2,\nu}(E)$. A direct calculation gives%
\begin{align}
A(E)  &  =\frac{\mu|\Gamma(\alpha_{+})|^{2}(2p/k_{0})^{2\mu}}{\Gamma^{2}%
(\beta_{+})\sin(2\pi\mu)}\left(  \mathrm{e}^{-\pi g_{1}/2p}\cos(2\pi
\mu)+\mathrm{e}^{\pi g_{1}/2p}\right)  +\tan\nu~,\nonumber\\
B(E)  &  =\frac{|\Gamma(\alpha_{+})|^{2}(2p/k_{0})^{2\mu}\mathrm{e}^{-\pi
g_{1}/2p}}{\Gamma^{2}(\beta)}>0~. \label{7.3.6c}%
\end{align}

For $E=-\tau^{2}<0$, $\tau>0$, $\lambda=2\tau$,$\ $the function $F_{2,\nu}(E)$
is real, therefore, $\sigma^{\prime}(E)$ can differ from zero only at the
discrete points $E_{n}(\nu)$ such that $F_{2,\nu}(E_{n}(\nu))=0$, or
$f_{2}(E_{n}(\nu))=-\tan\nu$, and we obtain that (derivatives with respect to
$E$ are denoted by primes in eq. (\ref{7.3.6cb}))%
\begin{align}
&  \sigma^{\prime}(E)=\sum_{n}\left[  -2\mu k_{0}F_{2,\nu}^{\prime}(E_{n}%
(\nu))\cos^{2}\nu\right]  ^{-1}\delta(E-E_{n}(\nu))~,\nonumber\\
&  F_{2,\nu}^{\prime}(E_{n}(\nu))=f_{2}^{\prime}(E_{n}(\nu))<0,\;\partial
_{\nu}E_{n}(\nu)=-\cos^{-2}\nu\left[  f_{2}^{\prime}(E_{n}(\nu))\right]
^{-1}>0~.\label{7.3.6cb}%
\end{align}

\textbf{I.} Let $g_{1}>0$

For $E=p^{2}>0$, $p>0$, the function $\sigma^{\prime}(E)\,$(\ref{7.3.6a}) is a
finite positive function.\ At $E=0$,\thinspace\ we have $B(0)=0$ and%
\[
\left.  A(0)\right|  _{\nu=\nu_{0}}=0,\;\tan\nu_{0}=-\Gamma(\beta_{-}%
)(g_{1}/k_{0})^{2\mu}\Gamma^{-1}(\beta_{+})~.
\]

It is easy to see that%
\begin{align*}
&  F_{2,\nu}(W)=\tan\nu-\tan\nu_{0}-\left(  2\mu k_{0}\cos^{2}\nu_{0}\right)
^{-1}\Psi^{-2}W+O(W),\;W\rightarrow0~,\\
&  \Psi=\frac{g_{1}(g_{1}/k_{0})^{-\mu}}{\mu\cos\nu_{0}}\sqrt{\frac
{3\Gamma(1+2\mu)}{2k_{0}(1+2\mu)\Gamma(2-2\mu)}}~.
\end{align*}
It follows that for $\nu\neq\nu_{0}$, $\,$the function $\sigma^{\prime}(E)$ is
finite at $E=0$. But for $\nu=\nu_{0}$ and for small $E$, we have:%
\[
\sigma^{\prime}(E)=-\frac{1}{\pi}\Psi^{2}\operatorname{Im}\left(  E+i0\right)
^{-1}+O(1)=\Psi^{2}\delta(E)+O(1)~,
\]
which means that there is the eigenvalue $E=0$ in the spectrum of the s.a.
Hamiltonian $\hat{H}_{2,\nu_{0}}$.

For $E=-\tau^{2}<0$, $\lambda=2\tau$, the function $f_{2}(E)$,%
\[
f_{2}(E)=\frac{\Gamma(\beta_{-})}{\Gamma(\beta_{+})}\frac{\Gamma(1/2+\mu
+g_{1}/2\tau)(2\tau/k_{0})^{2\mu}}{\Gamma(1/2-\mu+g_{1}/2\tau)}~,
\]
has the properties: $f_{2}(E)$ is smooth function for $E\in(-\infty,0)$,
$f_{2}(E)\rightarrow\infty$ as $E=-\infty$,$\;f_{2}(0)=-\tan\nu_{0}$. Because
$f_{2}^{\prime}(E_{\nu|n})<0$, see eq. (\ref{7.3.6cb}), the straight line
$f(E)=2\mu\tan\nu,E\in(-\infty,0]$, can intersect the plot of the function
$f_{2}(E)$ no more than once.

That is why the equation $F_{2,\nu}(E)=0$ has no solutions for $\nu\in(\nu
_{0},\pi/2)$ while for any fixed $\nu\in(-\pi/2,\nu_{0}]$, this equation has
only one solution $E^{\left(  -\right)  }\left(  \nu\right)  \in(-\infty,0],$
which increases monotonically from $-\infty$ to $0$ as $\nu$ changes from
$-\pi/2+0$ to $\nu_{0}$.

We thus obtain that the spectrum of $\hat{H}_{2,\nu},\left\vert \nu\right\vert
<\pi/2$, with $g_{1}>0$ is simple and given by
\begin{equation}
\mathrm{spec}\hat{H}_{2,\nu}=\left\{
\begin{array}
[c]{l}%
\mathbb{R}_{+}\cup\{E^{\left(  -\right)  }\left(  \nu\right)  \},\;\nu\in
(-\pi/2,\nu_{0}]\\
\mathbb{R}_{+},\;\nu\in(\nu_{0},\pi/2)\;\mathrm{or}\;\nu=\pm\pi/2
\end{array}
\right.  .\label{neg1a}%
\end{equation}
The generalized eigenfunctions%
\[
U_{E}(x)=\sqrt{\sigma^{\prime}(E)}u_{2,\nu}(x;E),\ E\geq0~,
\]
and (for $\nu\in(-\pi/2,\nu_{0}]$) the eigenfunction%
\[
U_{n}(x)=U(x)=\left[  -2\mu k_{0}F_{2,\nu}^{\prime}(E^{\left(  -\right)
}\left(  \nu\right)  ))\cos^{2}\nu\right]  ^{-1/2}u_{2,\nu}\left(
x;E^{\left(  -\right)  }\left(  \nu\right)  \right)
\]
of $\hat{H}_{2,\nu}$, form a complete orthonormalized systems in
$L^{2}(\mathbb{R}_{+})$.

\textbf{II.} Let $g_{1}<0$. Then:

For $E=p^{2}\geq0$, $p\geq0$, $\lambda=2p\mathrm{e}^{-i\pi/2}$, formulas
(\ref{7.3.6a}) and (\ref{7.3.6c}) hold true. Because the functions $A(E)$ and
$B(E)$ are finite at $E=0$ ($B(0)\neq0$), the function $\sigma^{\prime}(E)$
(\ref{7.3.6a}) is a finite positive function for $E\geq0$. This means that for
$E\geq0$, the spectra of s.a. Hamiltonians $\hat{H}_{2,\nu}$ are simple,
purely continuous, and given by \textrm{spec}$\hat{H}_{2,\nu}=\mathbb{R}_{+}$.

For $E=-\tau^{2}<0$, $\tau>0$, $\lambda=2\tau$, we have%
\[
f_{2}(E)=\frac{\Gamma(\beta_{-})}{\Gamma(\beta_{+})}\frac{\Gamma
(1/2+\mu-|g_{1}|/2\tau)(2\tau/k_{0})^{2\mu}}{\Gamma(1/2-\mu-|g_{1}|/2\tau)}~.
\]

It is easy to see that for fixed $\nu$, the spectrum is bounded from below and
the equation $F_{2,\nu}(E)=0$ has infinite number of solutions%
\begin{equation}
E_{n}\left(  \nu\right)  =-g_{1}^{2}/4n^{2}+O(n^{-3})~,\label{spect2}%
\end{equation}
asymptotically coinciding with (\ref{7.3c.3}) as $n\rightarrow\infty$.

We thus obtain that the spectrum of $\hat{H}_{2,\nu},\left|  \nu\right|
<\pi/2$, with $g_{1}<0$ is simple and given by $\mathrm{spec}\hat{H}_{2,\nu
}=\mathbb{R}_{+}\cup\{E_{n}\left(  \nu\right)  \}$. The corresponding
\ generalized eigenfunctions of the continuous spectrum%
\[
U_{E}(x)=\sqrt{\sigma^{\prime}(E)}u_{2,\nu}(x;E),\;E\geq0~,
\]
and eigenfunctions of the discrete spectrum%
\[
U_{n}(x)=\left[  -2\mu k_{0}F_{2,\nu}^{\prime}(E_{n}\left(  \nu\right)
)\cos^{2}\nu\right]  ^{-1/2}u_{2,\nu}(x;E_{n}\left(  \nu\right)
),\;E_{n}\left(  \nu\right)  <0~,
\]
of $\hat{H}_{2,\nu}$ form a complete orthonormalized system in $L^{2}%
(\mathbb{R}_{+})$.

It is possible to give a comparison description of the Hamiltonians $\hat
{H}_{2,\nu},|\nu|<\pi/2$ in more detail.

The function $f_{2}(E)$ has the properties: $f_{2}(E)\rightarrow\infty$ as
$E\rightarrow-\infty$; $f_{2}\left(  \mathcal{E}_{n}\pm0\right)  =\pm
\infty,\;n\in\mathbb{Z}_{+}$. Taking the third equality in (\ref{7.3.6cb})
into account, we can see that: in each energy interval $(\mathcal{E}%
_{n-1},\mathcal{E}_{n})$, $n\in\mathbb{Z}_{+}$, for a fixed $\nu\in(-\pi
/2,\pi/2)$, there are one discrete level $E_{n}(\nu)$ which increases
monotonically from $\mathcal{E}_{n-1}+0$ to $\mathcal{E}_{n}-0$ when $\nu$
changes from $\pi/2-0$ to $-\pi/2+0$ (we set $\mathcal{E}_{-1}=-\infty$). We
note that the relations%
\[
\lim_{\nu\rightarrow\pi/2}E_{n}\left(  \nu\right)  =\lim_{\nu\rightarrow
-\pi/2}E_{n+1}\left(  \nu\right)  =\mathcal{E}_{n},\;n\in\mathbb{Z}_{+}~,
\]
confirm the equivalence of s.a. extensions with parameters $\nu=-\pi/2$ and
$\nu=\pi/2$.

We note that it is possible to find the explicit expressions for spectrum,
spectral function, and the complete orthonormalized system of (generalized)
functions of the s.a. Hamiltonian for $\nu=0$. In this case, results are the
same as in the first range ($g_{2}\geq3/4$) with additional change
$\mu\rightarrow-\mu$. One can easily verify that such calculated spectrum
coincide with the spectrum $\{E_{n}\left(  0\right)  \}$.

It should be also pointed out that bound states exist even for the repulsive
potential, $g_{2}$,$g_{1}>0$, see the dashed line on the Figure 1.

\subsection{The third range $g_{2}=-1/4\ (\mu=0)$\label{SS3.3}}

The analysis in this section is similar to that in the previous one, a
peculiarity is that $\alpha_{+}=\alpha_{-}=\alpha=1/2+g_{1}/\lambda$,
$\beta_{+}=\beta_{-}=1$, $u_{1}\left(  x;W\right)  =u_{2}\left(  x;W\right)
$, and representation (\ref{7.3b.3}) of $\upsilon_{1}\left(  x;W\right)  \,$in
terms of $u_{1}$ and $u_{2}$ does not hold.\textrm{ }As the solutions of eq.
(\ref{7.3.1}) with $g_{2}=-1/4$, we therefore use the functions $u_{1}(x;W)$,
$u_{3}(x;W)$, and $\upsilon_{1}\left(  x;W\right)  $ respectively defined by%
\begin{align*}
&  u_{1}\left(  x;W\right)  =x^{1/2}\mathrm{e}^{-z/2}\Phi(\alpha,1;z)=\left.
u_{1}\left(  x;W\right)  \right\vert _{\lambda\rightarrow-\lambda}~,\\
&  u_{3}\left(  x;W\right)  =x^{1/2}\mathrm{e}^{-z/2}\frac{\partial}%
{\partial\mu}\left[  x^{\mu}\Phi(1/2+\mu+g_{1}/\lambda,1+2\mu;z)\right]
_{\mu=0}+u_{1}\left(  x;W\right)  \ln k_{0}~,\\
&  \upsilon_{1}\left(  x;W\right)  =x^{1/2}\mathrm{e}^{-z/2}\Psi
(\alpha,1;z)=\Gamma^{-1}(\alpha)\left[  \omega_{0}(W)u_{1}^{\left(  0\right)
}\left(  x;W\right)  -u_{3}\left(  x;W\right)  \right]  ~,\\
&  \omega_{0}(W)=2\psi(1)-\psi(\alpha)-\ln(\lambda/k_{0}),\ \ \alpha
=1/2+g_{1}/\lambda~,
\end{align*}
where $\psi(\alpha)=\Gamma^{\prime}(\alpha)/\Gamma(\alpha)$ and $k_{0}$ is a
constant. The functions $u_{1}\left(  x;W\right)  $ and $u_{3}\left(
x;W\right)  $ are real entire in $W$.

The asymptotic behavior of these functions at the origin and at infinity is
respectively as follows.

As $x\rightarrow0$, $z=\lambda x\rightarrow0$, we have%
\begin{align}
&  u_{1}(x;W)=k_{0}^{-1/2}u_{1\mathrm{as}}(x)+O(x^{3/2}),\ \ u_{1\mathrm{as}%
}(x)=(k_{0}x)^{1/2}~,\nonumber\\
&  u_{3}(x;W)=k_{0}^{-1/2}u_{3\mathrm{as}}(x)+O(x^{3/2}\ln
x),\ \ u_{3\mathrm{as}}(x)=(k_{0}x)^{1/2}\ln(k_{0}x)~,\nonumber\\
&  \upsilon_{1}(x;W)=k_{0}^{-1/2}\Gamma^{-1}(\alpha)\left[  \omega
_{0}(W)u_{1\mathrm{as}}\left(  x\right)  -u_{3\mathrm{as}}\left(  x\right)
\right]  +O(x^{3/2}\ln x)~.\label{7.3.8b}%
\end{align}

As $x\rightarrow\infty$, $\operatorname{Im}W>0$, we have%
\begin{align}
u_{1}(x;W) &  =\Gamma^{-1}(\alpha)\lambda^{\alpha-1}x^{g_{1}/\lambda
}\mathrm{e}^{z/2}\left[  1+O(x^{-1})\right]  \rightarrow\infty~,\nonumber\\
\upsilon_{1}(x;W) &  =\lambda^{-\alpha}x^{-g_{1}/\lambda}\mathrm{e}%
^{-z/2}\left[  1+O(x^{-1})\right]  \rightarrow0~.\label{7.3.10c}%
\end{align}

The functions $u_{1}$ and $u_{3}$ are linearly independent and form a
fundamental system of solutions of eq. (\ref{7.3.1}), as well as the functions
$u_{1}$ and $\upsilon_{1}$ for $\operatorname{Im}W\neq0$, see sec. \ref{S2},%
\[
\mathrm{Wr}\left(  u_{1},u_{3}\right)  =1,\ \mathrm{Wr}\left(  u_{1}%
,\upsilon_{1}\right)  =-\Gamma^{-1}(\alpha)~.
\]

We recall that, for $g_{2}=-1/4$, the deficiency indices of the initial
symmetric operator $\hat{H}$ are $m_{\pm}=1$, and therefore there exists a
one-parameter family of s.a. extensions of $\hat{H}$ with $g_{2}=-1/4$, see
sec. \ref{S2}.

To evaluate the asymmetry form in terms of a.b. coefficients, we need to
determine the asymptotics of functions $\psi_{\ast}$ belonging to the natural
domain $D_{\check{H}}^{\ast}(\mathbb{R}_{+})$ at the origin. To this end, we
use representation (\ref{7.3.4d}) of the general solution of eq.
(\ref{natDom}) with $W=0$\ where the natural substitutions $a_{2}%
u_{2}\rightarrow a_{2}u_{3}$ and $u_{2}/2\mu\rightarrow-u_{3}$ must be made.
Using the Cauchy-Bunyakovskii inequality for estimating the integral terms, we
obtain that the desired asymptotic as $x\rightarrow0$ is given by%
\begin{align*}
&  \psi_{\ast}(x)=a_{1}u_{1\mathrm{as}}(x)+a_{2}u_{3\mathrm{as}}%
(x)+O(x^{3/2}\ln x)~,\\
&  \psi_{\ast}^{\prime}(x)=a_{1}u_{1\mathrm{as}}^{\prime}(x)+a_{2}%
u_{3\mathrm{as}}^{\prime}(x)+O(x^{1/2}\ln x)~,
\end{align*}
and we find\footnote{This structure of $\Delta_{H^{+}}$ confirms the previos
assertion that the deficiency indices of $\hat{H}$ are $m_{\pm}=1$.}
$\Delta_{H^{+}}(\psi_{\ast})=k_{0}(\overline{a_{1}}a_{2}-\overline{a_{2}}%
a_{1})$, the coefficients $a_{1},a_{2}$ are just a.b. coefficients. The
requirement that $\Delta_{H^{+}}$ vanish results in the relation%
\[
a_{1}\cos\vartheta=a_{2}\sin\vartheta,\ \vartheta\in\mathbb{S}\left(
-\pi/2,\pi/2\right)  ~.
\]
This relation with fixed $\vartheta$ defines the domain of a possible
Hamiltonian as a s.a. restriction of $\hat{H}^{+}$, or a s.a. extension of
$\hat{H}$.

The final result is that: for $g_{2}=-1/4$ , there exists a family of s.a.
Hamiltonians $\hat{H}_{3,\vartheta}$ with the domains%
\[
D_{H_{3,\vartheta}}=\left\{  \psi:\psi\in D_{\check{H}}^{\ast}(\mathbb{R}%
_{+}),\ \psi\ \mathrm{satisfies}\ \text{(\ref{7.3.10a})}\right\}  ~,
\]
where (\ref{7.3.10a}) are the asymptotic s.a. boundary conditions at the
origin%
\begin{align}
&  \psi=C\psi_{3,\vartheta\mathrm{as}}(x)+O(x^{3/2}\ln x),\;\psi^{\prime
}=C\psi_{3,\vartheta\mathrm{as}}^{\prime}(x)+O(x^{1/2}\ln x)\ x\rightarrow
0~,\nonumber\\
&  \psi_{3,\vartheta\mathrm{as}}(x)=u_{1\mathrm{as}}(x)\sin\vartheta
+u_{3\mathrm{as}}(x)\cos\vartheta~.\label{7.3.10a}%
\end{align}

To evaluate the Green's function $G(x,y;W)$ for $\hat{H}_{3,\vartheta}$, we
take the representation (\ref{7.3.3a}) with $a_{1}=0$ for $\psi_{\ast}(x)$
belonging to $D_{H_{3,\vartheta}}\subset$ $D_{\check{H}}^{\ast}(\mathbb{R}%
_{+})$, boundary conditions (\ref{7.3.10a}) and asymptotics (\ref{7.3.8b})
then yield%
\[
a_{2}=-\Gamma^{2}(\alpha)\cos\vartheta\left[  \omega_{0}(W)\cos\vartheta
+\sin\vartheta\right]  ^{-1}\int_{0}^{\infty}\upsilon_{1}(x;W)\eta(x)dx\ .
\]
Using the representation%
\begin{align*}
&  \Gamma(\alpha)\upsilon_{1}=(\omega_{0}\sin\vartheta-\cos\vartheta
)u_{3,\vartheta}+(\omega_{0}\cos\vartheta+\sin\vartheta)\tilde{u}%
_{3,\vartheta}~,\\
&  u_{3,\vartheta}(x;W)=u_{1}(x;W)\sin\vartheta+u_{3}(x;W)\cos\vartheta~,\\
&  \tilde{u}_{3,\vartheta}(x;W)=u_{1}(x;W)\cos\vartheta-u_{3}(x;W)\sin
\vartheta~,
\end{align*}
where $u_{3,\vartheta}$\ and $\tilde{u}_{3,\vartheta}$ are solutions of eq.
(\ref{7.3.1}) real-entire in $W$, and $u_{3,\vartheta}$ satisfies boundary
condition (\ref{7.3.10a}), we find
\begin{align}
&  G(x,y;W)=\Omega(W)u_{3,\vartheta}(x;W)u_{3,\vartheta}(y;W)\nonumber\\
&  +\left\{
\begin{array}
[c]{c}%
\tilde{u}_{3,\vartheta}(x;W)u_{3,\vartheta}(y;W),\ x>y\\
u_{3,\vartheta}(x;W)\tilde{u}_{3,\vartheta}(y;W),\ x<y
\end{array}
\right.  ~,\nonumber\\
&  \Omega(W)=(\omega_{0}(W)\cos\vartheta+\sin\vartheta)^{-1}(\omega_{0}%
(W)\sin\vartheta-\cos\vartheta)~.\label{7.3.10b}%
\end{align}
We note that the second summand in $G(x,y;W)$ is real for real $W=E$.

It is easy to verify that the guiding functional given by (\ref{A.1.1}) with
$U=u_{3,\vartheta}$ satisfies the properties 1) and 3) cited in subsec.
\ref{SS3.1}. The proof that it satisfies the property 2) is identical to that
presented in subsec. \ref{SS3.2} for the second range $1>\mu>0$. It follows
that the spectra of $\hat{H}_{3,\vartheta}$ are simple.

The derivative of the spectral function is given by $\sigma^{\prime}%
(E)=\pi^{-1}\operatorname{Im}\left[  \Omega(E+i0)\right]  $.

We first consider the case $\vartheta=\pi/2$ where we have%
\begin{align*}
&  u_{3,\pi/2}(x;W)=u_{1}(x;W)~,\\
&  \sigma^{\prime}(E)=-\pi^{-1}\operatorname{Im}\Omega(E+i0),\ \Omega
(W)=\psi(\alpha)+\ln(\lambda/k_{0})~.
\end{align*}

For $E=p^{2}\geq0$, $p\geq0$, $\lambda=2p\mathrm{e}^{-i\pi/2}$, we find%
\[
\sigma^{\prime}(E)=\frac{1}{2}\left(  1-\tanh\frac{\pi g_{1}}{2p}\right)
\geq0~.
\]

For $E=-\tau^{2}<0$, $\tau>0$, $\lambda=2\tau$, and $g_{1}>0$, the function
$\Omega(E)$ is of the form%
\[
\Omega(E)=\psi(1/2+g_{1}/2\tau)+\ln(2\tau/k_{0})~,
\]
which implies that for $g_{1}>0$, there is no negative part of the spectrum.

For $E=-\tau^{2}<0$, $\tau>0$, $\lambda=2\tau$, and $g_{1}<0$, we have%
\[
\Omega(E)=\psi(1/2-|g_{1}|/2\tau)+\ln(2\tau/k_{0}),\ \operatorname{Im}%
\Omega(E)=\operatorname{Im}\psi(1/2-|g_{1}|/2\tau)~,
\]
which implies that there are discrete negative energy levels $\mathcal{E}_{n}$
in the spectrum,%
\begin{align*}
&  \mathcal{E}_{n}=-g_{1}^{2}(1+2n)^{-2},\ \tau_{n}=|g_{1}|(1+2n)^{-1}%
,\;n\in\mathbb{Z}_{+}\ \\
&  \sigma^{\prime}(E)=\sum_{n\in\mathbb{Z}_{+}}Q_{n}^{2}\delta(E-\mathcal{E}%
_{n}),\;Q_{n}=2|g_{1}|\left(  1+2n\right)  ^{-3/2}.
\end{align*}

It is easy to see that for the case of $\vartheta=-\pi/2$, we obtain the same
results for spectrum and eigenfunctions as it must be\textrm{.}

We thus obtain that for $g_{1}>0$, the spectrum of $\hat{H}_{3,\pm\pi/2}$ is
simple, continuous, and given by $\mathrm{spec}\hat{H}_{3,\pm\pi/2}%
=\mathbb{R}_{+}$, and a complete orthonormalized system in $L^{2}%
(\mathbb{R}_{+})$ of its generalized eigenfunctions consists of functions%
\[
U_{E}(x)=\sqrt{\sigma^{\prime}(E)}u_{1}(x;E),\ E\geq0\ .
\]

For $g_{1}<0$, the spectrum of $\hat{H}_{3,\pm\pi/2}$ is simple and given by
$\mathrm{spec}\hat{H}_{3,\pm\pi/2}=\mathbb{R}_{+}\cup\left\{  \mathcal{E}%
_{n},\ n\in\mathbb{Z}_{+}\right\}  $, and a complete orthonormalized system in
$L^{2}(\mathbb{R}_{+})$ of its (generalized) eigenfunctions consists of
functions%
\begin{align*}
&  U_{E}(x)=\sqrt{\sigma^{\prime}(E)}u_{1}(x;E),\ E\geq0\ ,\\
&  U_{n}(x)=2|g_{1}|\left(  1+2n\right)  ^{-3/2}u_{1}(x;\mathcal{E}%
_{n}),\ \mathcal{E}_{n}<0~.
\end{align*}

We note that the spectrum and eigenfunctions for $\hat{H}_{3,\pi/2}$ coincide
with those for $\hat{H}_{\mathfrak{e}}$ with $g_{2}\geq3/4$, if we set $\mu=0$
in the respective formulas in subsec. \ref{SS3.1}.

We now turn to the case $|\vartheta|<\pi/2$. In this case, $\sigma^{\prime
}(E)$ can be represented as%
\begin{align*}
&  \sigma^{\prime}(E)=(\pi\cos^{2}\vartheta)^{-1}\operatorname{Im}\left[
\omega_{3}(E+i0)\right]  ^{-1},\\
&  \omega_{3}(W)=\psi(\alpha)+\ln(\lambda/k_{0})-2\psi(1)-\tan\vartheta~.
\end{align*}

For $E=p^{2}\geq0$, $p\geq0$, $\lambda=2p\mathrm{e}^{-i\pi/2}$, and $g_{1}<0$,
we have%
\begin{equation}
\sigma^{\prime}(E)=\frac{B(E)}{\pi\cos^{2}\vartheta\lbrack A^{2}(E)+B^{2}%
(E)]}~,\label{7.3.11a}%
\end{equation}
where $\omega_{3}(E)=A(E)-iB(E)$. The function $B(E)$ can be explicitly
calculated:%
\begin{equation}
B(E)=\frac{\pi}{2}\left(  1-\tanh\frac{\pi g_{1}}{2\sqrt{E}}\right)
>0,\ \forall E\geq0~,\label{7.3.11b}%
\end{equation}
whence it follows that for all $E\geq0$, the spectrum of $\hat{H}%
_{3,\vartheta}$ is purely continuous.

For $E=p^{2}>0$, $p>0$, $\lambda=2p\mathrm{e}^{-i\pi/2}$, and $g_{1}>0$, the
spectral function is given by the same eqs. (\ref{7.3.11a}) and (\ref{7.3.11b}%
). But in this case, $B(0)=0$ and the limit $\lim_{W\rightarrow0}\omega
_{3}(W)$ must be carefully examined.

At small $W$, we have%
\[
\omega_{3}(W)=(\tan\vartheta_{0}-\tan\vartheta)-\left(  6g_{1}^{2}\right)
^{-1}W+O(W^{2}),\ \tan\vartheta_{0}=\ln(g_{1}/k_{0})-2\psi(1)~.
\]
For $\vartheta\neq\vartheta_{0}$, the function $\sigma^{\prime}(E)$ is finite
at $E=0$. But for $\vartheta=\vartheta_{0}$ and small $E$, we have%
\[
\sigma^{\prime}(E)=-\frac{6g_{1}^{2}}{\pi\cos^{2}\vartheta_{0}}%
\operatorname{Im}\left(  E+i0\right)  ^{-1}+O(1)=\frac{6g_{1}^{2}}{\cos
^{2}\vartheta_{0}}\delta(E)+O(1)~,
\]
which means that the spectrum of the Hamiltonian $\hat{H}_{3,\vartheta_{0}}$
contains an eigenvalue $E=0$.

For $E=-\tau^{2}<0$, $\tau>0$, $\lambda=2\tau$, the function $\omega_{3}(E)$
is real, therefore, $\sigma^{\prime}(E)$ can differ from zero only at
zero-points $E_{n}=E_{n}\left(  \vartheta\right)  $ of $\omega_{3}(E)$
($\omega_{3}(E_{n})=0$), which yields%
\begin{align}
&  \sigma^{\prime}(E)=\sum_{n}\left[  -k_{0}\omega_{3}^{\prime}(E_{n})\cos
^{2}\vartheta\right]  ^{-1}\delta(E-E_{n}),\ \omega_{3}^{\prime}%
(E_{n})<0~,\nonumber\\
&  \,\partial_{\vartheta}E_{n}\left(  \vartheta\right)  =\left[  \cos
^{2}\vartheta\omega_{3}^{\prime}(E_{n})\right]  ^{-1}<0~.\label{7.3.12}%
\end{align}

For $g_{1}>0$, we have%
\begin{align*}
\omega_{3}(E) &  =\psi(1/2+g_{1}/2\tau)+\ln(2\tau/g_{1})+\tan\vartheta
_{0}-\tan\vartheta~,\\
\omega_{3}(E) &  =(1/2)\ln|E|-\tan\vartheta+O(1),\ \ E\rightarrow-\infty~,\\
\omega_{3}(0) &  =\tan\vartheta_{0}-\tan\vartheta~.
\end{align*}

For $\vartheta<\vartheta_{0}$, the equation $\omega_{3}(E)=0$ has no solution,
whereas for $\vartheta\geq\vartheta_{0},$ it has only one solution $E^{\left(
-\right)  }\left(  \vartheta\right)  $. Because eq. (\ref{7.3.12}) holds for
$\partial_{\vartheta}E^{\left(  -\right)  }\left(  \vartheta\right)  $,
$E^{\left(  -\right)  }\left(  \vartheta\right)  $ increases from $-\infty$ to
$0$ when $\vartheta$ changes from $\pi/2-0$ to $\vartheta_{0}$.

For $g_{1}<0$, we have%
\begin{align*}
\omega_{3}(E) &  =\psi(1/2-|g_{1}|/2\tau)+\ln(2\tau/k_{0})-2\psi
(1)-\tan\vartheta~,\\
\omega_{3}(E) &  =(1/2)\ln|E|-\tan\vartheta+O(1),\ E\rightarrow-\infty~.
\end{align*}
It is easy to verify that the equation $\omega_{3}(E)=0$ has an infinite
number of solutions $E_{n},n\in\mathbb{Z}_{+}$, bounded from below and
asymptotically coinciding with (\ref{7.3c.3}) as $n\rightarrow\infty$,
$E_{n}=-g_{1}^{2}/4n^{2}+O(n^{-3})$.

We thus obtain that for $g_{1}>0$, the spectrum of $\hat{H}_{3,\vartheta}$ is
simple and given by $\mathrm{spec}\hat{H}_{3,\vartheta}=\mathbb{R}_{+}%
\cup\left\{  E^{\left(  -\right)  }\left(  \vartheta\right)  \right\}  $ and a
complete orthonormalized system in $L^{2}(\mathbb{R}_{+})$ of its
(generalized) eigenfunctions consists of functions%
\begin{align*}
&  U_{E}(x)=\sqrt{\sigma^{\prime}(E)}u_{3,\vartheta}(x;E),\ E\geq0\ ,\\
&  U(x)=\left[  -k_{0}\cos^{2}\vartheta\omega_{3}^{\prime}(E^{\left(
-\right)  }\left(  \vartheta\right)  )\right]  ^{-1/2}u_{3,\vartheta
}(x;E^{\left(  -\right)  }\left(  \vartheta\right)  )~,
\end{align*}
(the eigenvalue $E^{\left(  -\right)  }\left(  \vartheta\right)  $ exists, and
therefore $E^{\left(  -\right)  }\left(  \vartheta\right)  $ and the
corresponding eigenfunction $U(x)$ enter the inversion formulas only if
$\vartheta\geq\vartheta_{0}$); for $g_{1}<0$, the spectrum of $\hat
{H}_{3,\vartheta}$ is simple and given by $\mathrm{spec}\hat{H}_{3,\vartheta
}=\mathbb{R}_{+}\cup\left\{  E_{n}\right\}  $ and a complete orthonormalized
system in $L^{2}(\mathbb{R}_{+})$ of its (generalized) eigenfunctions consists
of functions%
\begin{align*}
&  U_{E}(x)=\sqrt{\sigma^{\prime}(E)}u_{3,\vartheta}(x;E),\ E\geq0\ ,\\
&  U_{n}(x)=\left[  -k_{0}\cos^{2}\vartheta\omega_{3}^{\prime}(E_{n})\right]
^{-1/2}u_{3,\vartheta}(x;E_{n}),\ E_{n}<0~.
\end{align*}

It is possible to describe the discrete spectrum for $|\vartheta|<\pi/2$ and
$g_{1}<0$ in more details. To this end, we represent the equation $\omega
_{3}(E)=0$ in the equivalent form%
\[
f_{3}(E)=\tan\vartheta,\;f_{3}(E)=\psi(1/2-|g_{1}|/2\tau)+\ln(2\tau
/k_{0})-2\psi(1)~.
\]
Then we have%
\[
f(-\infty)=\infty,\ \ f\left(  \mathcal{E}_{n}\pm0\right)  =\pm\infty
,\ n\in\mathbb{Z}_{+}.
\]
Because eq. (\ref{7.3.12}) holds, we can see that in each interval
$(\mathcal{E}_{n},\mathcal{E}_{n+1})$, $n\in\{-1\}\cup\mathbb{Z}_{+}$, there
is one discrete eigenvalue $E_{n}$ and $E_{n}$ increases monotonically from
$\mathcal{E}_{n}+0$ to $\mathcal{E}_{n+1}-0$ when $\vartheta$ changes from
$\pi/2-0$ to $-\pi/2+0$ (we set $\mathcal{E}_{-1}=-\infty$). We note the
relations%
\[
\lim_{\vartheta\rightarrow-\pi/2}E_{n-1}(\vartheta)=\lim_{\vartheta
\rightarrow\pi/2}E_{n}(\vartheta)=\mathcal{E}_{n}~.
\]

\subsection{The fourth range $g_{2}<-1/4$ $\left(  \mu=i\varkappa
,\varkappa>0\right)  $\label{SS3.4}}

The analysis in this section is completely similar to that in Section
\ref{SS3.2} (although the results for the spectrum differ drastically). We
therefore briefly outline basic points.

According to Section (\ref{S2}), the deficiency indices of the initial
symmetric operator $\hat{H}$ with $g_{2}<-1/4$ are $m_{\pm}=1$, and therefore
there exists a one-parameter family of\ its s.a. extensions.

To evaluate the asymmetry form $\Delta_{H^{+}}$,\ we determine the asymptotics
of functions $\psi_{\ast}$ belonging to $D_{\check{H}}^{\ast}(\mathbb{R}_{+})$
at the origin using representation (\ref{7.3.4d}) with $\mu=i\varkappa$ of the
general solution of eq. (\ref{natDom}) with $W=0$ and estimating the integral
terms by means of the Cauchy-Bunyakovskii inequality, which yields%
\begin{align}
&  \psi_{\ast}(x)=a_{1}u_{1\mathrm{as}}(x)+a_{2}u_{2\mathrm{as}}%
(x)+O(x^{3/2}),\ x\rightarrow0~,\nonumber\\
&  \psi_{\ast}^{\prime}(x)=a_{1}u_{1\mathrm{as}}^{\prime}(x)+a_{2}%
u_{2\mathrm{as}}^{\prime}(x)+O(x^{1/2}),\ x\rightarrow0~,\nonumber\\
&  u_{1\mathrm{as}}(x)=(k_{0}x)^{1/2+i\varkappa},\ u_{2\mathrm{as}}%
(x)=(k_{0}x)^{1/2-i\varkappa}=\overline{u_{1\mathrm{as}}(x)}~,\label{new}%
\end{align}
and we find\footnote{This structure of $\Delta_{H^{+}}$ confirms that the
deficiency indices of $\hat{H}$ are $m_{\pm}=1$.} $\Delta_{H^{+}}(\psi_{\ast
})=-2i\varkappa(\overline{a_{1}}a_{1}-\overline{a_{2}}a_{2})$. The requirement
that $\Delta_{H^{+}}$ vanish results in the relation $a_{1}=\mathrm{e}%
^{2i\theta}a_{2}$,$\ \theta\in\mathbb{S}\left(  0,\pi\right)  $ defining the
domains of possible s.a. Hamiltonians.

The final result is that: for each $g_{2}$ in the range $g_{2}<-1/4$, there
exists a family of s.a. Hamiltonians $\hat{H}_{4,\theta}$ with the domains%
\[
D_{H_{4,\theta}}=\left\{  \psi:\psi\in D_{\check{H}}^{\ast}(\mathbb{R}%
_{+}),\ \psi\ \mathrm{satisfies\ }\text{(\ref{7.3.13a})}\right\}  ~,
\]
where (\ref{7.3.13a}) are the asymptotic s.a. boundary conditions at the
origin%
\begin{align}
&  \psi=C\psi_{4\mathrm{as}}(x)+O(x^{3/2}),\;\psi^{\prime}=C\psi
_{4\mathrm{as}}^{\prime}(x)+O(x^{1/2}),\ x\rightarrow0~,\nonumber\\
&  \psi_{4\mathrm{as}}(x)=\mathrm{e}^{i\theta}u_{1\mathrm{as}}(x)+\mathrm{e}%
^{-i\theta}u_{2\mathrm{as}}(x)=\overline{\psi_{4\mathrm{as}}(x)}%
~.\label{7.3.13a}%
\end{align}

To evaluate the Green's function $G(x,y;W)$ for $\hat{H}_{4,\theta}$, we use
representation (\ref{7.3.3a}) with $a_{1}=0$ for $\psi_{\ast}(x)$ belonging to
$D_{H_{4,\theta}}\subset$ $D_{\check{H}}^{\ast}(\mathbb{R}_{+})$, boundary
conditions (\ref{7.3.6d}) and asymptotics (\ref{new}) then yield%
\begin{align*}
- &  a_{2}=\frac{2i\varkappa(\lambda k_{0})^{-i\chi}\mathrm{e}^{-i\theta}%
}{\omega(W)\omega_{4,\theta}(W)}\int_{0}^{\infty}\upsilon_{1}(x;W)\eta
(x)dx,\ \omega_{4,\theta}(W)=a(W)+b(W)~,\\
&  a(W)=\mathrm{e}^{i\theta}\frac{\Gamma(\beta)(\lambda/k_{0})^{-i\varkappa}%
}{\Gamma(\alpha)},\ b(W)=\mathrm{e}^{-i\theta}\frac{\Gamma(\beta_{-}%
)(\lambda/k_{0})^{i\varkappa}}{\Gamma(\alpha_{-})}\ .
\end{align*}
Using the representation%
\[
\upsilon_{1}(x;W)=-\frac{(\lambda/k_{0})^{i\varkappa}k_{0}^{-1/2+i\varkappa}%
}{4\varkappa}\left[  i\tilde{\omega}_{4,\theta}(W)u_{4,\theta}(x;W)+\omega
_{4,\theta}(W)\tilde{u}_{4,\theta}(x;W)\right]  ~,
\]
where%
\begin{align*}
&  \tilde{\omega}_{4,\theta}(W)=a(W)-b(W),\ \tilde{u}_{4,\theta}%
(x;W)=i[\mathrm{e}^{-i\theta}k_{0}^{1/2-i\varkappa}u_{2}(x;W)-\mathrm{e}%
^{i\theta}k_{0}^{1/2+i\varkappa}u_{1}(x;W)]~,\\
&  u_{4,\theta}(x;W)=\mathrm{e}^{i\theta}k_{0}^{1/2+i\varkappa}u_{1}%
(x;W)+\mathrm{e}^{-i\theta}k_{0}^{1/2-i\varkappa}u_{2}(x;W)~,
\end{align*}
where $u_{4,\theta}$ and $\tilde{u}_{4,\theta}$ are solutions of eq.
(\ref{7.3.1}) real-entire in $W$, and $u_{4,\theta}$ satisfies boundary
conditions (\ref{7.3.13a}), we find
\begin{align*}
&  G(x,y;W)=\Omega(W)u_{4,\theta}(x;W)u_{4,\theta}(y;W)\\
&  -\frac{1}{4\varkappa k_{0}}\left\{
\begin{array}
[c]{c}%
\tilde{u}_{4,\theta}(x;W)u_{4,\theta}(y;W),\ x>y\\
u_{4,\theta}(x;W)\tilde{u}_{4,\theta}(y;W),\ x<y
\end{array}
\right.  ,\ \Omega(W)=-\frac{i}{4\varkappa k_{0}}\frac{\tilde{\omega
}_{4,\theta}(W)}{\omega_{4,\theta}(W)}~,
\end{align*}
the second summand in $G(x,y;W)$ is real for real $W=E$.

It is easy to verify that the guiding functional given by (\ref{A.1.1}) with
$U=u_{4,\theta}$ satisfies the properties 1)- 3) cited in subsec. \ref{SS3.1},
whence it follows that the spectra of $\hat{H}_{4,\theta}$ are simple.

The derivative of the spectral function is given by $\sigma^{\prime}%
(E)=\pi^{-1}\operatorname{Im}\Omega(E+i0).$

For $E=p^{2}\geq0$, $p\geq0$, $\lambda=2p\mathrm{e}^{-i\pi/2}$, and $g_{1}<0$,
we have%
\begin{align}
&  \sigma^{\prime}(E)=\pi^{-1}\operatorname{Im}\Omega(E)=\frac{\left(
4\pi\varkappa k_{0}\right)  ^{-1}\left(  1-|D(E)|^{2}\right)  }%
{(1+D(E))(1+\overline{D}(E))}~,\label{7.3.15.a}\\
&  D(E)=\frac{a(E)}{b(E)}=\frac{\mathrm{e}^{-2i\theta}\Gamma(\beta
)\Gamma(\alpha_{-})\mathrm{e}^{2i\varkappa\ln(k_{0}/2p)}\mathrm{e}%
^{-\pi\varkappa}}{\Gamma(\beta_{-})\Gamma(\alpha)}~.\nonumber
\end{align}
Because%
\begin{equation}
|D(E)|^{2}=\frac{1+\mathrm{e}^{-2\pi\varkappa}\mathrm{e}^{-\pi g_{1}/p}%
}{1+\mathrm{e}^{2\pi\varkappa}\mathrm{e}^{-\pi g_{1}/p}}<1,\ \forall
p\geq0~,\label{7.3.15b}%
\end{equation}
\textrm{spec}$\hat{H}_{4,\theta}=\mathbb{R}_{+}$ and is simple.

For $E=p^{2}>0$, $p>0$, $\lambda=2p\mathrm{e}^{-i\pi/2}$, and $g_{1}>0$
expressions (\ref{7.3.15.a}) and (\ref{7.3.15b}) for $\sigma^{\prime}(E)$ hold
true. But in this case, we have $|D(0)|=1$ and must carefully examine the
limit $\lim_{W\rightarrow0}\Omega(W)$.

It is easy to see that for small $W$, we have the representation%
\begin{align*}
&  \Omega(W)=-\frac{i}{4\varkappa k_{0}}\frac{1+\mathrm{e}^{2i(\theta
_{0}-\theta)}}{[1-\mathrm{e}^{2i(\theta_{0}-\theta)}]+iW/A}+O(1),\ A=\frac
{3g_{1}{}^{2}}{\varkappa(1+4\varkappa^{2})}~,\\
&  \theta_{0}=\varphi-\pi\lbrack\varphi/\pi],\ \varphi=\varkappa\ln
(g_{1}/k_{0})-\theta_{\Gamma}+\pi/2,\ \theta_{\Gamma}=\frac{1}{2i}\ln
\frac{\Gamma(\beta)}{\Gamma(\beta_{-})}~,
\end{align*}
where $[\varphi/\pi]$ is the entire part of $\varphi/\pi$. For $\theta
\neq\theta_{0}$, the function $\sigma^{\prime}(E)$ is finite at $E=0$. But for
$\theta=\theta_{0}$, we find%
\[
\sigma^{\prime}(E+0)=-\pi^{-1}\left(  A/2\varkappa k_{0}\right)
\operatorname{Im}\left(  E+i0\right)  ^{-1}+O(1)=\left(  A/2\varkappa
k_{0}\right)  \delta(E)+O(1)~,
\]
which means that the spectrum of the Hamiltonian $\hat{H}_{4,\theta_{0}}$ with
$g_{1}>0$ contains the eigenvalue $E=0$.

For $E=-\tau^{2}<0$, $\tau>0$, $\lambda=2\tau$, the function $\Omega$ \ can be
represented as%
\[
\Omega(E)=\pi\tan\Theta(E),\ \Theta(E)=\theta+\theta_{\Gamma}-\theta_{\Gamma
}(E)+\varkappa\ln(k_{0}/2\tau)~,
\]
where%
\begin{align*}
&  \theta_{\Gamma}(E)=\frac{1}{2i}\left[  \ln\Gamma(1/2+g_{1}/2\tau
+i\varkappa)-\ln\Gamma(1/2+g_{1}/2\tau-i\varkappa)\right]  \\
&  =\left\{
\begin{array}
[c]{l}%
\left\{
\begin{array}
[c]{l}%
-\pi|g_{1}|/2\tau+\varkappa\ln(|g_{1}|/2\tau)+O(1),\ g_{1}<0\\
\varkappa\ln(g_{1}/2\tau)+O(\tau),\ g_{1}>0
\end{array}
\right.  ,\ E\rightarrow0\\
\theta_{\Gamma}(-\infty)=\frac{1}{2i}\ln\frac{\Gamma(1/2+i\varkappa)}%
{\Gamma(1/2-i\varkappa)}+O(1/\tau),\ E\rightarrow-\infty
\end{array}
\right.  .
\end{align*}
The asymptotic behavior of $\Theta(E)$ at the origin and at minus infinity is
given by%
\[
\Theta(E)=\left\{
\begin{array}
[c]{c}%
\left\{
\begin{array}
[c]{l}%
\pi|g_{1}|/2\tau+O(1),\ g_{1}<0\\
\theta+\theta_{\Gamma}+\varkappa\ln(k_{0}/g_{1})+O(\tau),\ g_{1}>0
\end{array}
\right.  ,\ E\rightarrow0\\
\theta+\theta_{\Gamma}-\theta_{\Gamma}(-\infty)+\varkappa\ln(k_{0}%
/2\tau)+O(1/\tau),\ E\rightarrow-\infty
\end{array}
\right.  .
\]

Because $\Omega(E)$ is a real function for $E<0$, $\sigma^{\prime}\left(
E\right)  $ can differ from zero only at the points $E_{n}=$ $E_{n}(\theta)$
where $\Theta(E_{n})=\pi/2+\pi n$, $n\in\mathbb{Z}$\textrm{, }which yields%
\[
\sigma^{\prime}(E)=\sum_{n}Q_{n}^{2}\delta(E-E_{n}),\ Q_{n}=\left[  4\varkappa
k_{0}\Theta^{\prime}(E_{n})\right]  ^{-1/2},\;\Theta^{\prime}(E_{n})>0.
\]

We can obtain an additional information about the discrete spectrum of
$\hat{H}_{4,\theta}$. Representing the equation $\Theta(E_{n})=\pi/2+\pi n$,
$n\in\mathbb{Z}$, in the equivalent form%
\begin{align*}
&  f_{4}(E_{n})=\pi/2+\pi(n-\theta/\pi),\;f_{4}(E)=\theta_{\Gamma}%
-\theta_{\Gamma}(E)+\varkappa\ln(k_{0}/2\tau)~,\\
&  \ \partial_{\theta}E_{n}(\theta)=-\left[  f_{4}^{\prime}(E_{n}%
(\theta))\right]  ^{-1}=-\left[  \Theta^{\prime}(E_{n}(\theta))\right]
^{-1}<0~,
\end{align*}
we can see that the following assertions hold.

\begin{enumerate}
\item[a)] The eigenvalue $E_{n}(\theta)$ with fixed $n$ decreases
monotonically from $E_{n}\left(  0\right)  $\ to $E_{n}\left(  \pi\right)  -0$
when $\theta$ changes from $0$ to $\pi-0$. In particular, we have
$E_{n-1}(\theta)<E_{n}(\theta)$, $\forall n$.

\item[b)] For any $g_{1}$, the spectrum is unbounded from below:
$E_{n}\rightarrow-\infty$ as $n\rightarrow-\infty$.

\item[c)] For any $\theta$, the negative part of the spectrum is of the form
$E_{n}=-k_{0}^{2}m^{2}\mathrm{e}^{2\pi|n|/\varkappa}(1+O(1/n))$ as
$n\rightarrow-\infty$, where $m=m(g_{1},g_{2},\theta)$ is a scale factor, and
asymptotically (as\textrm{ }$n\rightarrow-\infty$) coincides with the negative
part of the spectrum in the Calogero model with coupling constant $g_{2}$
under an appropriate identification of scale factors.

\item[d)] For $g_{1}<0$, the discrete part of the spectrum has an accumulation
point $E=0$. More specifically, the spectrum is of the form $E_{n}=-g_{1}%
^{2}/4n^{2}+O(1/n^{3})$ as $n\rightarrow\infty$ (as in all the previous ranges
of the parameter $g_{2}$) and asymptotically coincides with the spectrum for
$g_{2}=0$, see below.

\item[e)] For $g_{1}>0$, the discrete spectrum has no finite accumulation
points. In particular, possible values of $n$ are restricted from above,
$n\leq n_{\max}$, where%
\[
n_{\max}=\left\{
\begin{array}
[c]{l}%
f_{4}(0)/\pi-1/2\ \mathrm{if}\ f_{4}(0)/\pi-1/2\ \mathrm{is\ integer}\\
\left[  f_{4}(0)/\pi+1/2\right]  \ \mathrm{if\ }f_{4}(0)/\pi-1/2>[f_{4}%
(0)/\pi-1/2]
\end{array}
\right.  ,
\]
and the level $E=0$ is present in the spectrum for $\theta=\theta_{0}$ only.
\end{enumerate}

The final result is as follows: the spectrum of $\hat{H}_{4,\theta}$ is simple
and given by $\mathrm{spec}\hat{H}_{4,\theta}=\mathbb{R}_{+}\cup\{E_{n}%
\leq0\},\ -\infty<n<n_{\max}$, where\ $n_{\max}<\infty$ for $g_{1}>0$ and
$n_{\max}=\infty$ for $g_{1}<0$, and the set of the corresponding
(generalized) eigenfunctions%
\[
U_{E}(x)=\sqrt{\sigma^{\prime}(E)}u_{4,\theta}(x;E),\ E\geq0;\ \ U_{n}%
(x)=Q_{n}u_{4,\theta}(x;E_{n}),\ E_{n}\leq0,
\]
form a complete orthonormalized system in $L^{2}\left(  \mathbb{R}_{+}\right)
$.

\subsection{The fifth range $g_{2}=0$ ($\mu=1/2$)\label{SS3.5}}

The analysis in this section is similar to that in subsec. \ref{SS3.2}. A
peculiarity is that the function $u_{2}$ is not defined for $\mu=1/2$, and we
therefore use the following solutions of eq. (\ref{7.3.1}):%
\begin{align*}
u_{1}(x;W) &  =x\mathrm{e}^{-z/2}\Phi(\alpha_{1/2},2;z),\ u_{5}(x;W)=\tilde
{u}_{5}(x;W)-g_{1}\ln k_{0}u_{1}(x;W)~,\\
\upsilon_{1}(x;W) &  =x\mathrm{e}^{-z/2}\Psi(\alpha_{1/2},2;z)=\Gamma
^{-1}(\alpha_{1/2})\left[  \omega_{1/2}(W)u_{1}(x;W)+u_{5}(x;W)\right]  ~,
\end{align*}
where%
\begin{align*}
&  \alpha_{1/2}=1+g_{1}/\lambda,\ \\
&  \tilde{u}_{5}(x;W)=\mathrm{e}^{-z/2}x^{1/2}\left[  x^{-\mu}\Phi(\alpha
_{-},\beta_{-};z)+g_{1}\Gamma(\beta_{-})x^{\mu}\Phi(\alpha_{+},\beta
_{+};z)\right]  _{\mu\rightarrow1/2}~,\\
&  \omega_{1/2}(W)=g_{1}\mathbf{C+}g_{1}\left[  \psi(\alpha_{1/2})+\ln
(\lambda/k_{0})\right]  -g_{1}-\lambda/2~,
\end{align*}
$\mathbf{C}$ is the Euler constant. The asymptotics of these functions at the
origin and at infinity are respectively as follows.

As $x\rightarrow0$, $z=\lambda x\rightarrow0$, we have%
\begin{align}
&  u_{1}(x;W)=k_{0}^{-1}u_{1\mathrm{as}}(x)+O(x^{2}),\ u_{5}%
(x;W)=u_{5\mathrm{as}}(x)+O(x^{2}\ln x)~,\nonumber\\
&  \upsilon_{1}(x;W)=\Gamma^{-1}(\alpha_{1/2})\left[  k_{0}^{-1}\omega
_{1/2}(W)u_{1\mathrm{as}}(x)+u_{5\mathrm{as}}(x)\right]  +O(x^{2}\ln
x)~,\nonumber\\
&  u_{1\mathrm{as}}(x)=k_{0}x,\ u_{5\mathrm{as}}(x)=1+g_{1}x\ln(k_{0}%
x)+\mathbf{C}g_{1}x~.\label{7.3.15d}%
\end{align}
As $x\rightarrow\infty$, $\operatorname{Im}W>0$, we have%
\begin{align*}
u_{1}(x;W) &  =\Gamma^{-1}(\alpha_{1/2})\lambda^{-1+g_{1}/\lambda}%
x^{+g_{1}/\lambda}\mathrm{e}^{z/2}(1+O(x^{-1}))\rightarrow\infty~,\\
\upsilon_{1}(x;W) &  =\lambda^{-g_{1}/\lambda}x^{-g_{1}/\lambda}%
\mathrm{e}^{-z/2}(1+O(x^{-1}))\rightarrow0~.
\end{align*}
The functions $u_{1}\left(  x;W\right)  $ and $u_{5}\left(  x;W\right)  $ are
real-entire in $W$. These functions form a fundamental system of solutions of
eq. (\ref{7.3.1}), the same holds for the functions $u_{1},\upsilon_{1}$ for
$\operatorname{Im}W\neq0$, see subsec. \ref{SS3.2},%
\[
\mathrm{Wr}\left(  u_{1},u_{5}\right)  =-1,\ \mathrm{Wr}\left(  u_{1}%
,\upsilon_{1}\right)  =-1/\Gamma(\alpha_{1/2})=-\omega(W)~.
\]

As we know from subsec. \ref{SS3.2}, for $g_{2}<-1/4$, the deficiency indices
of the initial symmetric operator $\hat{H}$ are $m_{\pm}=1$, and therefore
there exists a one-parameter family of\ its s.a. extensions.

For evaluating the asymmetry form $\Delta_{H^{+}}$,\ we determine the
asymptotics of functions $\psi_{\ast}$, belonging to $D_{\check{H}}^{\ast
}(\mathbb{R}_{+})$, at the origin using representation (\ref{7.3.4d}) of the
general solution of eq. (\ref{natDom}) with $W=0$, where the natural
substitutions\ $a_{2}u_{2}\rightarrow a_{2}u_{5}$ and $u_{2}/2\mu\rightarrow
u_{5}$ must be made, and estimating the integral terms by means of the
Cauchy-Bunyakovskii inequality, which yields%
\begin{align}
\psi_{\ast}(x) &  =a_{1}u_{1\mathrm{as}}(x)+a_{2}u_{5\mathrm{as}}%
(x)+O(x^{3/2})~,\nonumber\\
\psi_{\ast}^{\prime}(x) &  =a_{1}u_{1\mathrm{as}}^{\prime}(x)+a_{2}%
u_{5\mathrm{as}}^{\prime}(x)+O(x^{1/2})~,\label{7.3.16b}%
\end{align}
and we find\footnote{This structure of $\Delta_{H^{+}}$ confirms that the
deficiency indices of $\hat{H}$ are $m_{\pm}=1$.} $\Delta_{H^{+}}(\psi_{\ast
})=-k_{0}(\overline{a_{1}}a_{2}-\overline{a_{2}}a_{1})$. The requirement that
$\Delta_{H^{+}}$ vanish results in the relation $a_{1}\cos\epsilon=a_{2}%
\sin\epsilon$,\ $\epsilon\in\mathbb{S}\left(  -\pi/2,\pi/2\right)  .$

The final result is that for $g_{2}=0$, there exists a family of s.a.
Hamiltonians $\hat{H}_{5,\epsilon}$ with the domains%
\[
D_{H_{5,\epsilon}}=\left\{  \psi:\psi\in D_{\check{H}}^{\ast}(\mathbb{R}%
_{+}),\ \psi\ \mathrm{satisfies\ }\text{(\ref{7.3.17a})}\right\}  ~,
\]
where (\ref{7.3.17a}) are the asymptotic s.a. boundary conditions at the
origin%
\begin{align}
&  \psi=C\psi_{5,\epsilon\mathrm{as}}(x)+O(x^{3/2}),\;\psi^{\prime}%
=C\psi_{5,\epsilon\mathrm{as}}^{\prime}(x)+O(x^{1/2}),\ x\rightarrow
0~,\nonumber\\
&  \psi_{5,\epsilon\mathrm{as}}(x)=u_{1\mathrm{as}}(k_{0}x)\sin\epsilon
+u_{5\mathrm{as}}(x)\cos\epsilon~.\label{7.3.17a}%
\end{align}

To find the Green's function $G(x,y;W)$ for $\hat{H}_{5,\epsilon}$, we use
representation (\ref{7.3.3a}) with $a_{1}=0$ for $\psi_{\ast}(x)$ belonging to
$D_{H_{4,\theta}}\subset$ $D_{\check{H}}^{\ast}(\mathbb{R}_{+})$, boundary
conditions (\ref{7.3.17a}) and asymptotics (\ref{7.3.15d}) then yield%
\[
a_{2}=-\frac{\Gamma^{2}(\alpha_{1/2})\cos\epsilon}{\omega_{1/2}(W)\cos
\epsilon-k_{0}\sin\epsilon}\int_{0}^{\infty}\upsilon_{1}(x;W)\eta(x)dx~.
\]
Using the representation%
\begin{align*}
&  k_{0}\Gamma(\alpha_{1/2})\upsilon_{1}(x;W)=(\omega_{1/2}(W)\cos
\epsilon-k_{0}\sin\epsilon)\tilde{u}_{5,\epsilon}(x;W)\\
&  +(\omega_{1/2}(W)\sin\epsilon+k_{0}\cos\epsilon)u_{5,\epsilon}(x;W)~,\\
&  u_{5,\epsilon}(x;W)=k_{0}u_{1}(x;W)\sin\epsilon+u_{5}(x;W)\cos\epsilon~,\\
&  \tilde{u}_{5,\epsilon}(x;W)=k_{0}u_{1}(x;W)\cos\epsilon-u_{5}%
(x;W)\sin\epsilon~,
\end{align*}
where $u_{5,\epsilon}(x;W)$ and $\tilde{u}_{5,\epsilon}(x;W)$ are solutions of
eq. (\ref{7.3.1}) real-entire in $W$ and $u_{5,\epsilon}(x;W)$ satisfies
boundary conditions (\ref{7.3.17a}), we find%
\begin{align*}
&  G(x,y;W)=\frac{1}{k_{0}}\left[  \Omega(W)u_{5,\epsilon}(x;W)u_{5,\epsilon
}(y;W)-\left\{
\begin{array}
[c]{c}%
\tilde{u}_{5,\epsilon}(x;W)u_{5,\epsilon}(y;W),\ x>y\\
u_{5,\epsilon}(x;W)\tilde{u}_{5,\epsilon}(y;W),\ x<y
\end{array}
\right.  \right]  ,\\
&  \Omega(W)=\left[  k_{0}\sin\epsilon-\omega_{1/2}(W)\cos\epsilon\right]
^{-1}\left[  \omega_{1/2}(W)\sin\epsilon+k_{0}\cos\epsilon\right]  ~,
\end{align*}
the second summand in $G(x,y;W)$ is real for real $W=E$.

It is easy to verify that the guiding functional given by (\ref{A.1.1}) with
$U=u_{5,\epsilon}$ satisfies the properties 1)-3) cited in subsec.
\ref{SS3.1}, whence it follows that the spectra of $\hat{H}_{5,\epsilon}$ are simple.

The derivative of the spectral function is given by $\sigma^{^{\prime}%
}(E)=\left(  \pi k_{0}\right)  ^{-1}\operatorname{Im}\Omega(E+i0)$.

We first consider the case of $\epsilon=\pi/2$ where we have $u_{5,\pi
/2}(x;W)=k_{0}u_{1}(x;W)$ and%
\begin{align*}
&  \sigma^{\prime}(E)=\left(  \pi k_{0}^{2}\right)  ^{-1}\operatorname{Im}%
\tilde{\Omega}(E+i0)~,\\
&  \tilde{\Omega}(W)=g_{1}\psi(\alpha_{1/2})+g_{1}\ln(\lambda/k_{0}%
)-\lambda/2~.
\end{align*}

For $E=p^{2}\geq0$, $p\geq0$, $\lambda=2p\mathrm{e}^{-i\pi/2}$, we have%
\[
\sigma^{\prime}(E)=\frac{|g_{1}|\mathrm{e}^{-\pi g_{1}/2p}}{2k_{0}^{2}%
\sinh(\pi|g_{1}|/2p)}\geq0~.
\]

For $E=-\tau^{2}<0$, $\tau>0$, $\lambda=2\tau$, and $g_{1}>0$, $\alpha
_{1/2}=1+g_{1}/2\tau$, the function $\tilde{\Omega}(E)$ is finite and real,
whence it follows that there are no negative spectrum points.

For $E=-\tau^{2}<0$, $\tau>0$, $\lambda=2\tau$,$\,$and $g_{1}<0$,
$\alpha_{1/2}=1-|g_{1}|/2\tau$, we have%
\begin{align*}
&  \sigma^{\prime}(E)=-\left(  \pi k_{0}^{2}\right)  ^{-1}|g_{1}%
|\operatorname{Im}\left.  \psi(\alpha)\right\vert _{W=E+i0}=\sum
_{n\in\mathbb{Z}_{+}}Q_{n}^{2}\delta(E-\mathcal{E}_{n})~,\\
&  \mathcal{E}_{n}=-\frac{g_{1}^{2}}{(2+2n)^{2}},\;Q_{n}=\frac{2}{k_{0}%
}\left(  \frac{|g_{1}|}{2+2n}\right)  ^{3/2}.
\end{align*}

It is easy to see that for the case of $\epsilon=-\pi/2$, we obtain the same
results for spectrum and eigenfunctions as it must be\textrm{.}

We thus obtain that for $g_{1}>0$, the spectrum of $\hat{H}_{5,\pi/2}$ is
simple, continuous, and given by$\,\mathrm{spec}\hat{H}_{5,\pm\pi
/2}=\mathbb{R}_{+}$ and the set of generalized eigenfunctions $U_{E}%
(x)=\sqrt{\sigma^{\prime}(E)}u_{5,\pi/2}(x;E)$,$\ E\geq0$, form a complete
orthonormalized system in $L^{2}\left(  \mathbb{R}_{+}\right)  $.

For $g_{1}<0$, the spectrum of $\hat{H}_{5,\pm\pi/2}$ is simple and given by
$\mathrm{spec}\hat{H}_{5,\pm\pi/2}=\mathbb{R}_{+}\cup\left\{  \mathcal{E}%
_{n},\ n\in\mathbb{Z}_{+}\right\}  $ and the set of (generalized)
eigenfunctions%
\begin{align*}
&  U_{E}(x)=\sqrt{\sigma^{\prime}(E)}u_{5,\pi/2}(x;E),\ E\geq0~,\\
&  U_{n}(x)=\frac{2}{k_{0}}\left(  \frac{|g_{1}|}{2+2n}\right)  ^{3/2}%
u_{5,\pi/2}(x;\mathcal{E}_{n})~,
\end{align*}
form a complete orthonormalized system in $L^{2}\left(  \mathbb{R}_{+}\right)
$.

We now turn to the case $|\epsilon|<\pi/2$ where we have%
\[
\sigma^{\prime}(E)=\left(  \pi\cos^{2}\epsilon\right)  ^{-1}\operatorname{Im}%
\left[  \omega_{5}(E+i0)\right]  ^{-1},\ \ \omega_{5}(W)=k_{0}\tan
\epsilon-\omega_{1/2}(W)~.
\]

For $g_{1}<0$, $E=p^{2}\geq0$, $p\geq0$, $\lambda=2p\mathrm{e}^{-i\pi/2}$, we
obtain that%
\begin{equation}
\sigma^{^{\prime}}(E)=\left(  \pi\cos^{2}\epsilon\right)  ^{-1}%
\operatorname{Im}\omega_{5}^{-1}(E)=\frac{B(E)}{\pi\cos^{2}\epsilon\lbrack
A^{2}(E)+B^{2}(E)]}~, \label{7.3.19a}%
\end{equation}
where $\omega_{5}(E)=A(E)-iB(E)$. The function $B(E)$ is explicitly given by%
\begin{equation}
B(E)=\frac{\pi}{2}\frac{|g_{1}|\mathrm{e}^{-\pi g_{1}/2p}}{\sinh(\pi
|g_{1}|/2p)}>0,\ \forall p\geq0~. \label{7.3.19b}%
\end{equation}

It follows that for $g_{1}<0$,$\ E\geq0$, the spectrum of $\hat{H}%
_{5,\epsilon}$ is purely continuous.

For $g_{1}>0$, $E=p^{2}>0$, $p>0$, $\lambda=2p\mathrm{e}^{-i\pi/2}$, the
derivative of the spectral function is also given by eqs. (\ref{7.3.19a}) and
(\ref{7.3.19b}). But in this case, we have $B(0)=0$ and the limit
$\lim_{W\rightarrow0}\omega_{5}(W)$ has to be carefully examined. For small
$W$, we have%
\begin{align*}
&  \omega_{5}(W)=(\tan\epsilon-\tan\epsilon_{0})k_{0}-\frac{1}{3g_{1}%
}W+O(W^{2})~,\\
&  \tan\epsilon_{0}=(g_{1}/k_{0})\left[  \ln(g_{1}/k_{0})+\mathbf{C}-1\right]
~.
\end{align*}
For $\epsilon\neq\epsilon_{0}$, the function $\sigma^{\prime}(E)$ has a finite
limit as $E\rightarrow0$. But for $\epsilon=\epsilon_{0}$ and small $E$, we
have%
\[
\sigma^{\prime}(E)=-\frac{3g_{1}}{\pi\cos^{2}\epsilon_{0}}\operatorname{Im}%
\left(  E+i0\right)  ^{-1}+O\left(  1\right)  =\frac{3g_{1}}{\cos^{2}%
\epsilon_{0}}\delta(E)+O\left(  1\right)  ~,
\]
which means that the spectrum of the Hamiltonian $\hat{H}_{5,\epsilon_{0}}$
has an eigenvalue $E=0$.

For $E=-\tau^{2}<0$, $\tau>0$, $\lambda=2\tau$, the function $\omega_{5}(E)$
is real. Therefore, $\sigma^{\prime}(E)$ can differ from zero only at zero
points $E_{n}=E_{n}(\epsilon)$ of $\omega_{5}(E)$, and $\sigma^{\prime}(E)$ is
represented as%
\[
\sigma^{\prime}(E)=\sum_{n}\left[  -\omega_{5}^{\prime}(E_{n})\right]
^{-1}\delta(E-E_{n}),\ \omega_{5}(E_{n})=0,\ \omega_{5}^{\prime}(E_{n})<0~.
\]

For $g_{1}>0$, we have%
\begin{align*}
&  \omega_{5}(E)=-g_{1}\psi(1+g_{1}/2\tau)-g_{1}\ln(2\tau/g_{1})+\tau
+k_{0}(\tan\epsilon-\tan\epsilon_{0})~,\\
&  \omega_{5}(E)=\sqrt{|E|}-(g_{1}/2)\ln|E|+O(1),\ \ E\rightarrow
-\infty;\ \ \omega_{5}(0)=k_{0}(\tan\epsilon-\tan\epsilon_{0})~.
\end{align*}

For $\epsilon>\epsilon_{0},$ the equation $\omega_{5}(E)=0$ has no solution,
while for $\epsilon\in(-\pi/2,\epsilon_{0}]$ it has a unique solution
$E^{\left(  -\right)  }\left(  \epsilon\right)  $. It is easy to see that%
\[
\partial_{\epsilon}E^{\left(  -\right)  }\left(  \epsilon\right)
=-k_{0}[\omega_{5}^{\prime}\left(  E_{\epsilon}^{\left(  -\right)  }\right)
\cos^{2}\epsilon]^{-1}>0~,
\]
so that $E^{\left(  -\right)  }\left(  \epsilon\right)  $ increases
monotonically from $-\infty$ to $0$ when $\epsilon$ changes from $-\pi/2+0$ to
$\epsilon_{0}$.

For $g_{1}<0$, we have%
\begin{align*}
&  \omega_{5}(E)=|g_{1}|\psi(1/2-|g_{1}|/2\tau)+|g_{1}|\ln(2\tau/k_{0}%
)+\tau-\tilde{\epsilon}~,\\
&  \tilde{\epsilon}=g_{1}\mathbf{C}-g_{1}-k_{0}\tan\epsilon~\mathbf{.}%
\end{align*}

Representing the equation $\omega_{5}(E_{n})=0$ in the equivalent form%
\[
f_{5}(E_{n})=\tilde{\epsilon},\ f_{5}(E)=|g_{1}|\psi(1/2-|g_{1}|/2\tau
)+|g_{1}|\ln(2\tau/k_{0})+\tau~,
\]

we can see that:

\begin{enumerate}
\item[a)]
\[
f_{5}(E)\overset{E\rightarrow-\infty}{\longrightarrow}\infty,\ \ f_{5}\left(
\mathcal{E}_{n}\pm0\right)  =\pm\infty~,
\]
such that in each region of energy $(\mathcal{E}_{n},\mathcal{E}_{n+1})$,
$n\in(-1)\cup\mathbb{Z}_{+}$, the equation $\omega_{5}(E_{n})=0$ has one
solution $E_{n}(\epsilon)$ for any fixed $\epsilon$, $|\epsilon|<\pi/2$, and
$E_{n}(\epsilon)$ increases monotonically from $\mathcal{E}_{n}+0$ to
$\mathcal{E}_{n+1}-0$ as $\epsilon$ changes from $-\pi/2+0$ to $\pi/2-0$
(here, by the definition, $\mathcal{E}_{-1}=-\infty$).

\item[b)] For any fixed $\epsilon$, $E_{n}(\epsilon)=-g_{1}^{2}/4n^{2}%
+O(n^{-3})$ as $n\rightarrow\infty$, asymptotically coinciding with
(\ref{7.3c.3}).

\item[c)] The point $E=0$ is an accumulation point of discrete spectrum for
$g_{1}<0$.
\end{enumerate}

Note the relation%
\[
\lim_{\epsilon\rightarrow\pi/2}E_{n-1}(\epsilon)=\lim_{\epsilon\rightarrow
-\pi/2}E_{n}(\epsilon)=\mathcal{E}_{n},\;n\in\mathbb{Z}_{+}~.
\]

The above results can be briefly summarized as follows.

For $g_{1}<0$, the spectrum of $\hat{H}_{5,\epsilon}$ is simple and given by
$\mathrm{spec}\hat{H}_{5,\epsilon}=\mathbb{R}_{+}\cup\{E_{n}<0$,\ $n\in
(-1)\cup\mathbb{Z}_{+}\}$. The (generalized) eigenfunctions%
\begin{align*}
&  U_{E}(x)=\sqrt{\sigma^{^{\prime}}(E)}u_{5,\epsilon}(x;E),\ E\geq0~,\\
&  U_{n}(x)=\left[  -\omega_{5,\epsilon}^{\prime}(E_{n})\right]
^{-1/2}u_{5,\epsilon}(x;E_{n}),\ E_{n}<0,\;n\in(-1)\cup\mathbb{Z}_{+}~,
\end{align*}
form a complete orthonormalized system in $L^{2}\left(  \mathbb{R}_{+}\right)
$.

For $g_{1}>0$, the spectrum of $\hat{H}_{5,\epsilon}$ is simple and given by
$\mathrm{spec}\hat{H}_{5,\epsilon}=\mathbb{R}_{+}\cup\{E^{\left(  -\right)
}\left(  \epsilon\right)  \leq0\}$. For $\epsilon\in(-\pi/2,\epsilon_{0}]$ the
(generalized) eigenfunctions%
\[
U_{E}(x)=\sqrt{\sigma^{^{\prime}}(E)}u_{5,\epsilon}(x;E),\ E\geq
0,\ U(x)=\left[  -\omega_{5}^{\prime}(E^{\left(  -\right)  })\right]
^{-1/2}u_{5,\epsilon}\left(  x;E^{\left(  -\right)  }\right)
\]
form a complete orthonormalized system in $L^{2}\left(  \mathbb{R}_{+}\right)
$. For $\epsilon>\epsilon_{0}$, the spectrum has no negative eigenvalues.

We note that the above results (for spectrum and eigenfunctions) can be
extracted from the results in subsec. \ref{SS3.2}\ for the case $g_{2}\neq0$
($\mu\neq1/2$).

\section{Some concluding remarks \label{S4}}

We would like to finish our consideration with a remark about the Kratzer
potential \cite{Kra20} mentioned in the Introduction. This potential
corresponds to a particular case of parameters $g_{2}>0$ and $g_{1}<0$. It is
drown by the thick line in the graph of Figure 1. As was already said, the
Kratzer potential is extensively used to describe the molecular structure and
interactions \cite{RoyBe70}. In such cases, the Kratzer potential appears in
the radial part of the Schr\"{o}dinger equation (\ref{radial}) and has the
form:%
\begin{equation}
V\left(  x\right)  =-2D_{e}\left(  \frac{a}{x}-\frac{1}{2}\frac{a^{2}}{x^{2}%
}\right)  ~, \label{kratzer}%
\end{equation}
where $D_{e}$ is the dissociation energy and $a$ is the equilibrium
inter-nuclear separation. As $x$ goes to zero, $V\left(  x\right)  $ goes to
infinity, describing the internuclear repulsion and, as $x$ goes to infinity,
$V\left(  x\right)  $ goes to zero, describing the decompositions of
molecules. Putting the potential (\ref{kratzer}) in the radial equation
(\ref{radial}) and comparing with the Schr\"{o}dinger equation (\ref{7.3.1}),
we have the following identification:%
\[
g_{1}=-\frac{4m}{\hbar^{2}}D_{e}a~,\ g_{2}=\frac{2m}{\hbar^{2}}D_{e}%
a^{2}+l\left(  l+1\right)  \ .
\]
We can now calculate the value of $g_{2}$ for real diatomic molecules. Using
data from \cite{KarPo70}, even for $l=0$, we have $g_{2}=4.53\times10^{4}$ for
CO. The parameter $g_{2}$ is of the same order for molecules of NO, O$_{2}$,
I$_{2}$, and H$_{2}$. Thus, we can see that for the realistic Kratzer
potentials, the corresponding radial equations have always $g_{2}>3/4$. Thus,
the corresponding radial problem belongs to the first range described in
subsec. \ref{SS3.1}. In this case, there exist only one s.a. radial
Hamiltonian defined on the natural domain (\ref{1}), functions from this
domain have asymptotics (\ref{natdom2}).

\begin{acknowledgement}
M.C.B. thanks FAPESP; D.M.G. thanks FAPESP and CNPq for permanent support;
I.T. thanks RFBR Grand 08--01-00737; I.T. and B.V. thank Grand LSS-1615.2008.2
for partial support.
\end{acknowledgement}


\begin{thebibliography}{99}                                                                                               %


\bibitem {Kra20}A. Kratzer, \emph{Die ultraroten Rotationsspektren der
Halogenwasserstoffe}, Z. Phys \textbf{3}, 289 (1920)

\bibitem {Bau53}E.C. Baughan, \emph{Comments on the thermochemistry of the
elements of Groups IVB and IV}, Quart. Rev. \textbf{7}, 103 (1953)

\bibitem {Fue26}E. Fues, \emph{Ueber die Bestimmung der mittleren W\"{a}rme
der Luft}, Ann. Physik \textbf{80}, 376 (1926)

\bibitem {BalMi90}C.J. Ballhausen and M. Gajhede, \emph{The tunnel effect and
scattering by a negative Kratzer potential}, Chemical Physics Letters
\textbf{165}, Issue 5, 449 (1990)

\bibitem {BayBoC07}O. Bayrak I. Boztosun and H. Ciftci, \emph{Exact Analytical
Solutions to the Kratzer Potential by the Asymptotic Iteration Method}, Int J
Quantum Chem \textbf{107}, 540 (2007)

\bibitem {Shor31}G.H. Shortley, \emph{The inverse-cube force in quantum
mechanics}, Phys. Rev. \textbf{38}, 120 (1931)

\bibitem {Scar58}S.L. Scarf, \emph{Discrete States for Singular Potential
Problems}, Phys.Rev. \textbf{109}, 2170 (1958)

\bibitem {LanLi77}L.D. Landau and E.M. Lifshitz, \emph{Quantum Mechanics}%
\textit{ }(Pergamon Press 1977).

\bibitem {Eli62}M.A. Eliashevich, \emph{Atomic and Molecular Spectroscopy}
(State Physical and Mathematical Publishing, Moscow, 1962)

\bibitem {Flu94}S. Fl\"{u}gge, \emph{Practical Quantum Mechanics}; Vol I
(Springer: Berlin, 1994)

\bibitem {Mas49}H.S.W. Massey, \emph{Theory of Atomic Collisions} (Clarendon
Press, Oxford, 1949)

\bibitem {VorGiT107}B.L. Voronov, D.M. Gitman, and I.V. Tyutin,
\emph{Constructing Quantum Observables and Self-Adjoint Extensions of
Symmetric Operators. I}, Russian Physics Journal \textbf{50/}1, 1 (2007);
\emph{Constructing Quantum Observables and Self-Adjoint Extensions of
Symmetric Operators. II. Differential Operators}, Russian Physics Journal
\textbf{50/}9, 853\textbf{\ }(2007); \emph{Constructing quantum observables
and self --adjoint extensions of symmetric operators. III. Self--adjoint
boundary conditions,} Russian Physics Journ. \textbf{51}/2, 115 (2008)

\bibitem {Neu29}J. von Neumann, \emph{Zur Algebra der Funktionaloperationen
und Theorie der normalen Operatoren}, Math. Ann. \textbf{102}, 370 (1929)

\bibitem {Krein46}M.T. Krein, \emph{One general method of decompositions of
positively defined kernals in elementar products}, DAN USSR \textbf{53}, 3
(1946); \emph{Hermitian operators with guiding functionals}, Zbirnik Prazc'
Institutu Matematiki, AN USR No.10, 83 (1948)

\bibitem {Naima69}M.A. Naimark, \emph{Linear Differential Operators} (Nauka,
Moscow 1969); N.I. Akhiezer and I.M. Glazman, \emph{Theory of Linear Operators
in Hilbert Space} (Pitman, Boston 1981)

\bibitem {RoyBe70}R.J. Le Roy, R.B. Bernstain, \emph{Dissociation Energy and
Long-Range Potential of Diatomic Molecules from Vibrational Spacings of Higher
Levels}, J. Chem. Phys. \textbf{52}, 3869 (1970)

\bibitem {VigSi05}J. Vigo-Aguiar, T. E. Simos, \emph{Review of multistep
methods for the numerical solution of the radial Schr\"{o}dinger equation},
Int J Quantum Chem \textbf{103}, 278 (2005)

\bibitem {BatEr53}H. Bateman and A. Erdelyi, \emph{Higher Transcedental
Functions} (Mc GRAW-HILL, New York, 1953)

\bibitem {GraRy71}Gradshtein I.S., Ryzhik N.M. \emph{Tables of Integrals,
Sums, Series and Products}. Nauka, Moscow, 1971.

\bibitem {KarPo70}M. Karplus, R.N. Porter,  \emph{Atoms and Molecules} (WA
Benjamin: Menlo Park, CA, 1970)

\bibitem {ReeSi72}M. Reed and B. Simon, \emph{Methods of Modern Mathematical
Physics vol 2 Harmonic Analysis. Self-adjointness} (New York: Academic,1972)

\bibitem {BagGi90}V.G. Bagrov and D.M. Gitman, \emph{Exact Solutions of
Relativistic Wave Equations}, (Kluwer, Dordrecht, Boston, London 1990)

\bibitem {Neu32}J. von Neumann, \emph{Mathematical Foundations of Quantum
Mechanics} (Princeton University Press, Princeton 1955)

\bibitem {VorGiT307}B.L. Voronov, D.M. Gitman and I.V. Tyutin, \emph{The Dirac
Hamiltonian with a superstrong Coulomb field}, Theor. Math. Phys. \textbf{150}
34 (2007); \emph{Self-adjoint extensions and spectral analysis in Calogero
problem}, J. Phys. A \textbf{43}, 145205 (2010)

\bibitem {GitTySV09}D.M. Gitman, I.V. Tyutin, A.G. Smirnov, and B.L. Voronov,
\emph{Self-adjoint Schr\"{o}dinger and Dirac operators with Aharonov-Bohm and
magnetic-solenoid fields}, arXiv:0911.0946 [quant-ph] (2009)
\end{thebibliography}
\end{document}